\def\PUBLICPREPRINT{1}

\documentclass{article}

\usepackage{iclr2027_conference,times}
\usepackage{amsmath,amssymb,amsthm}
\usepackage{booktabs}
\usepackage{graphicx}
\usepackage{tabularx}
\usepackage{float}
\usepackage{tikz}
\usetikzlibrary{arrows.meta,calc,positioning,decorations.pathreplacing}
\usepackage{pgfplots}
\pgfplotsset{compat=1.18}
\usepackage{microtype}
\usepackage[hidelinks]{hyperref}
\usepackage{url}

\newif\ifpublicpreprint
\ifdefined\PUBLICPREPRINT
  \publicpreprinttrue
\else
  \publicpreprintfalse
\fi

\newtheorem{theorem}{Theorem}
\newtheorem{lemma}{Lemma}
\newtheorem{proposition}{Proposition}
\newtheorem{corollary}{Corollary}

\newcommand{\one}{\mathbf 1}
\newcommand{\ind}[1]{\mathbf 1\{#1\}}
\newcommand{\norm}[1]{\left\lVert #1 \right\rVert}
\newcommand{\pos}[1]{\left[#1\right]_+}

\title{CFR without Unbiasedness:\\
Debit-Budgeted Public Feedback}

\ifpublicpreprint
  \author{
    Jiaxing Guo\\
    Imperial College London\\
    \texttt{henry.guo23@imperial.ac.uk}
    \And
    Lei Ye\\
    Imperial College London\\
    \texttt{kevin.ye23@imperial.ac.uk}
  }
  \iclrfinalcopy
\else
  \author{Anonymous authors}
\fi

\begin{document}

\maketitle
\ifpublicpreprint
  \lhead{Preprint}
\fi

\begin{abstract}
At a finite public-chance cut, partial counterfactual regret minimization (CFR)
can evaluate successive outcome batches at learner states changed by earlier
batches.  Balanced coverage then lacks the conditional unbiasedness used by
Monte Carlo CFR analyses.  We introduce \emph{target debit}, the one-sided
cumulative underestimation needed to transfer regret on delivered feedback to
the dense target game.  Our central theorem bounds weighted exploitability by
delivered-row regret, target debit, and a native alternating-update correction.
This yields Debit-Budgeted CFR (DB-CFR), which selects a public plan from
completed history, executes it once, and settles its debit afterward.  Complete
coverage has zero debit and supplies exact recovery.  With at most $\ell$
unsettled blocks, hidden debit is bounded by the sum of the $\ell$ largest
predictable block envelopes; for fixed $\ell$ and polynomial output weights,
the resulting $O(T^{-1})$ term preserves the additive-RM/RM+ rate.  A finite
converse gives equal-metadata plans with charges $0$ and $11/8$, proving that a
nontrivial strict-budget decision requires semantic information or bounded
post-execution settlement.  In two released heads-up no-limit hold'em turn
endgames, all 42 registered shallow comparisons favor partial coverage before
complete coverage overtakes it between 32 and 64 rounds.  In public Leduc, a
lagged rule fixed on four development orders beats the best
matched-recurrence-work fixed width on all ten unseen orders (median NashConv
gain $0.0456$) without future candidate replay.

\end{abstract}

\section{Introduction}
\label{sec:introduction}

Counterfactual regret minimization (CFR) computes approximate equilibria in
large imperfect-information games by alternating counterfactual evaluation
with local regret updates \citep{zinkevich2007cfr}.  Public chance often
creates a finite cut of outcomes observed by both players.  Evaluating every
outcome gives an exact update; evaluating a smaller batch lowers the work per
update and permits more frequent learner transitions.  Public Chance Sampling
and Monte Carlo CFR (MCCFR) obtain guarantees by making each sampled update a
conditionally valid estimator of its target
\citep{lanctot2009mccfr,johanson2012pcs,farina2020stochastic}.

Many high-throughput solvers use a different pattern.  They process a fixed
public order without replacement, update the learner after each batch, and
continue the order across the epoch.  Every outcome appears equally often,
but later batches are evaluated at profiles produced by earlier feedback.
The delivered row is therefore coupled to its own history.  Exact counts at
the end of an epoch do not make the intermediate rows conditionally unbiased,
and regret on those rows need not control exploitability in the full game.

The two-outcome case isolates the issue.  Let $X_A(\sigma)$ and
$X_B(\sigma)$ be centered public components at profile $\sigma$.  Processing
$A$ and then $B$ delivers $2X_A(\sigma_t)$, updates the strategy, and then
delivers $2X_B(\sigma_{t+1})$.  Relative to dense feedback at both origins,
the epoch error is
\[
 [X_A(\sigma_t)-X_A(\sigma_{t+1})]
 -[X_B(\sigma_t)-X_B(\sigma_{t+1})].
\]
Coverage cancels static selection error, but not the interaction between
component identity and learner motion.  We call the resulting one-sided
full-game correction the \emph{target debit};
Figure~\ref{fig:two-outcome-mechanism} illustrates the mechanism.

\begin{figure}[t]
\centering
\begin{tikzpicture}[
  x=1mm, y=1mm,
  every node/.style={font=\footnotesize},
  evt/.style={draw, rounded corners=1pt, align=left, inner sep=2.2pt},
  err/.style={align=left, text=black!75},
  lab/.style={font=\footnotesize\itshape}
]
\draw[-{Stealth[length=2mm]}] (0,0) -- (122,0);
\node[anchor=west] at (122.5,0) {\footnotesize round};

\node[evt, anchor=south west] (e1) at (4,6)
  {event $t$ at profile $\sigma_t$:\\
   delivered $2X_{A,t}$\quad target $X_{A,t}+X_{B,t}$};
\node[err, anchor=north west] at (4,4.5)
  {error $\;+\,(X_{A,t}-X_{B,t})$};
\draw (12,0) -- (12,1.5);

\node[evt, anchor=south west] (e2) at (64,6)
  {event $t{+}1$ at profile $\sigma_{t+1}$:\\
   delivered $2X_{B,t+1}$\quad target $X_{A,t+1}+X_{B,t+1}$};
\node[err, anchor=north west] at (68,4.5)
  {error $\;-\,(X_{A,t+1}-X_{B,t+1})$};
\draw (72,0) -- (72,1.5);

\draw[-{Stealth[length=1.8mm]}, thick]
  (e1.north east) to[out=55,in=125,looseness=1.35]
  node[midway, above=2.2mm] {\footnotesize RM/RM+ update}
  (e2.north west);

\draw[decorate, decoration={brace, mirror, amplitude=2.5mm}]
  (4,-6) -- (118,-6);
\node[align=center, text width=126mm, anchor=north] at (61,-9.5)
  {epoch error $=(X_{A,t}-X_{A,t+1})-(X_{B,t}-X_{B,t+1})$.  Coverage cancels
   the selection coefficients; only \emph{component motion} remains.  At
   complete coverage, both outcomes are consumed at one profile and the error
   is zero.};
\end{tikzpicture}
\caption{Balanced selection cancels static component error.  Learner updates
between partial batches leave target debit; complete coverage evaluates all
components at one origin and has zero debit.}
\label{fig:two-outcome-mechanism}
\end{figure}

Our central result turns this error into a decision account.  For each local
row, only cumulative coordinates on which delivered feedback
\emph{underestimates} the dense target can increase target regret.
Theorem~\ref{thm:weighted-target-debit} combines this one-sided row transfer
with the native alternating CFR reduction.  It bounds weighted exploitability
by delivered-row regret, target debit, and a native alternating correction.

Debit-Budgeted CFR (DB-CFR) uses this account causally.  At an aligned block
boundary, it selects a width and order from completed history, executes that
plan once, and settles its charge afterward.  Partial execution continues
while the settled account is solvent and at most $\ell$ audits remain
unsettled; otherwise the method executes complete coverage, whose debit is
exactly zero.  Theorem~\ref{thm:debit-controller} shows that hidden debit is at
most the sum of the $\ell$ largest predictable block envelopes.  For fixed
blocks and polynomial output weights, a fixed audit cap contributes only
$O(T^{-1})$ and preserves the additive-RM/RM+ rate.  Thus target validity does
not require either per-event unbiasedness or future candidate replay.

The settled charge must contain semantic information.  Our finite converse compares
two four-round plans with the same public counts, work, task graph, prefix
discrepancy, and scalar strategy-path length.  Their exact controller charges
are $0$ and $11/8$.  Any sound rule based only on those shared summaries must
charge both by at least $11/8$ and therefore rejects the safe plan at budget
one.  A strict-budget admission rule limited to the same summaries must also
reject both before execution.  DB-CFR resolves this information boundary by
settling semantic debit after the selected block and reserving for the bounded
outstanding account.

The empirical program first establishes the decision geometry, then changes
the learner path.  Released HUNL turn endgames show that partial width dominates
at shallow budgets, that order changes quality at fixed coverage, and that
complete coverage overtakes every partial width between 32 and 64 full-cut
budgets.  In public Leduc, widths 2, 5, and 10 occupy one work--debit frontier.
A lagged completed-history rule fixed on four development orders then beats the
best fixed width at no greater installed recurrence work on all ten unseen
orders, with median NashConv improvement 0.0456 and no future candidate
replay.  A separate HUNL resource audit measures the exact selected-path
certificate rather than treating it as free work.

Our contributions are threefold.  \textbf{(i)} We establish a weighted,
one-sided target-debit theorem for deterministic partial public feedback under
native alternating CFR.  \textbf{(ii)} We introduce causal DB-CFR and prove
that bounded settlement removes future candidate replay while preserving the
additive-RM/RM+ rate; strict separations distinguish its charge from symmetric
and metadata-only alternatives.  \textbf{(iii)} We provide two-domain
evidence for the public width/order problem, finite trace certificates, and a
held-out dynamic-width consequence.

\section{Deterministic Public Feedback as a Control Problem}
\label{sec:background}

\paragraph{Game, local rows, and output.}
Fix a finite two-player zero-sum perfect-recall game with player-zero payoff
$u_0=u=-u_1$.  Let $x$ and $y$ be realization plans and define
\begin{equation}
\operatorname{expl}(x,y)=\tfrac12\left[
\max_{x'}u(x',y)-\min_{y'}u(x,y')\right].
\label{eq:gap-definition}
\end{equation}
For information set $I$ owned by player $i$, write $\rho=(i,I)$ for its local
row and $p_{\rho,t}$ for its behavior at round $t$.  If
$v_{\rho,t}$ is the counterfactual action-value vector, its centered feedback
is $v_{\rho,t}-\langle p_{\rho,t},v_{\rho,t}\rangle\one$.  Additive signed
regret matching (RM) accumulates centered rows; additive RM+ clips the updated
state coordinatewise at zero \citep{tammelin2014cfrplus}.  In both cases,
regret matching normalizes the positive part of the state and uses the
uniform distribution when its positive mass is zero, matching the solver
implementation.  We use the native
alternating recurrence
$(x_t,y_t)\to(x_{t+1},y_t)\to(x_{t+1},y_{t+1})$ and predetermined
nonnegative nondecreasing output weights $w_t$.  Put
$W_T=\sum_{t=1}^T w_t>0$ and denote the weighted pre/pre average by
$(\bar x_T^w,\bar y_T^w)$.

\paragraph{Finite additive public cut.}
Let $J$ contain $N$ semantic public labels.  For row $\rho$, label $j$, and
the immutable consuming profile at event $t$, let
$X_{\rho,t,j}$ be the centered contribution of label $j$.  A width-$B$ plan,
where $B$ divides $N$, selects $S_t\subseteq J$ with $|S_t|=B$ and delivers
\begin{equation}
g_{\rho,t}=\sum_{j\in J}X_{\rho,t,j},\qquad
h_{\rho,t}=\frac{N}{B}\sum_{j\in S_t}X_{\rho,t,j}.
\label{eq:public-rows}
\end{equation}
Here $g_{\rho,t}$ is the dense full-cut target and $h_{\rho,t}$ is the row
consumed by the learner.  Exactly evaluated terms outside the public cut are
shared by both rows and cancel from their difference.  The appendix extends
the notation to finitely many nonnested cuts and proves the additive
decomposition from terminal-history linearity.

Define the row error
\begin{equation}
\epsilon_{\rho,t}=h_{\rho,t}-g_{\rho,t}
\label{eq:row-error}
\end{equation}
and, for any finite vector $z$, the one-sided underestimation gauge
\begin{equation}
\psi(z)=\pos{\max_a\{-z(a)\}}=\norm{[-z]_+}_\infty.
\label{eq:psi}
\end{equation}
If $z$ is coordinatewise nonnegative, delivered cumulative feedback dominates
the target and $\psi(z)=0$, even when $z\ne0$.  During player zero's update,
define the target-payoff decrease
\begin{equation}
d_t=\pos{u(x_t,y_t)-u(x_{t+1},y_t)}.
\label{eq:target-decrease}
\end{equation}
The public part of a block's exact target debit is the sum of the row gauges;
the alternating part is the weighted sum of $d_t$.

The common feedback regimes use different guarantees.  Complete coverage is
exact at every origin, and independent public sampling is target valid through
conditional expectation.  A persistent partial order instead balances counts
across events with different learner origins; its connection to
$g_{\rho,t}$ must therefore be established pathwise.

\paragraph{Control decision and event-faithful execution.}
At a completed block boundary, a candidate plan $a$ fixes a common number $m$
of alternating rounds, a width $B$, a balanced public order, and an admissible
physical executor.  The planner selects a plan from completed history and a
settled debit account.  The selected plan executes once; its target-debit
charge may be certified before execution or settled by an audit after the
block.  Thus width and order are learner decisions, while processor placement
is permitted only when it preserves the declared event record.

All selected components of one learner event must use one immutable profile
and join in a deterministic canonical order.  All rows owned by the updating
player read the resulting common delivered payoff functional and update
simultaneously in exactly one learner transition.  We call this contract
\emph{event-faithful}.  Physical
parallelism inside the join may change runtime but not the declared learner
trajectory.  Complete coverage is the plan $B=N$, for which
$h_{\rho,t}=g_{\rho,t}$ at every event.

\section{Target Debit and Exact-Fallback Control}
\label{sec:theory}

We first transfer regret from delivered rows to the dense target game, then
turn the resulting correction into a causal plan constraint.

Define delivered weighted positive local regret
\begin{equation}
\widehat{\mathcal R}_T^w=
\sum_\rho\pos{\max_a\sum_{t=1}^T w_t h_{\rho,t}(a)}.
\label{eq:delivered-positive-regret}
\end{equation}
Let
$\Omega_0=\max_{x,x',y}\{u(x,y)-u(x',y)\}$, which is finite on the
realization-plan polytopes.

\begin{theorem}[Weighted target-debit transfer]
\label{thm:weighted-target-debit}
For every finite event-faithful execution under the declared additive public
cut contract,
\begin{equation}
\boxed{
\operatorname{expl}(\bar x_T^w,\bar y_T^w)
\le\frac{1}{2W_T}\left[
\widehat{\mathcal R}_T^w
+\sum_\rho\psi\!\left(\sum_{t=1}^T w_t\epsilon_{\rho,t}\right)
+\Omega_0w_T+\sum_{t=1}^T w_td_t
\right].}
\label{eq:weighted-target-debit}
\end{equation}
\end{theorem}

\paragraph{Proof.}
Put $H_\rho=\sum_t w_th_{\rho,t}$ and
$E_\rho=\sum_t w_t\epsilon_{\rho,t}$, so the cumulative target row is
$H_\rho-E_\rho$.  Rowwise,
\begin{equation}
\pos{\max_a(H_\rho(a)-E_\rho(a))}
\le \pos{\max_a H_\rho(a)}+\psi(E_\rho).
\label{eq:one-sided-row-transfer}
\end{equation}
Indeed, for every action $a$,
$H_\rho(a)-E_\rho(a)\le\max_bH_\rho(b)+\max_b[-E_\rho(b)]$;
taking the maximum and then the positive part proves the display.  Summing it
over the finite row registry bounds the two players' exact target local
regrets by $\widehat{\mathcal R}_T^w+\sum_\rho\psi(E_\rho)$.

For completeness, fix a valid comparator for one player.  The perfect-recall
policy-difference identity writes its weighted external regret as a sum over
owned information sets of the comparator's own reach, which lies in $[0,1]$,
times a comparator mixture of the cumulative centered counterfactual row.
That mixture is at most the row maximum, and replacing that maximum by its
positive part preserves the bound.  Applying this argument at
$(x_t,y_t)$ for player zero and at $(x_{t+1},y_t)$ for player one, then taking
the two comparator suprema, gives the native alternating local-to-external
reduction.  Bilinearity of $u$ identifies its weighted comparator sums with
the gap of the normalized weighted realization-plan averages.  The only
unmatched term is
\begin{align}
C_T^w={}&\max_y\sum_t w_tu(x_{t+1}-x_t,y)
-\sum_t w_tu(x_{t+1}-x_t,y_t). \label{eq:weighted-bridge}
\end{align}
Summation by parts gives
\[
\sum_t w_t(x_{t+1}-x_t)=w_Tx_{T+1}-w_1x_1
-\sum_{t=2}^T(w_t-w_{t-1})x_t.
\]
The negative endpoint coefficient has magnitude $w_1$, while the increment
coefficients have total $w_T-w_1$; together their magnitude is $w_T$.
Because $W_T>0$ and the clock is nonnegative and nondecreasing, $w_T>0$.
The negative terms divided by $w_T$ therefore form a convex combination
$x^-$ of feasible realization plans, and, for every $y$,
\[
\sum_t w_tu(x_{t+1}-x_t,y)
=w_T\{u(x_{T+1},y)-u(x^-,y)\}\le \Omega_0w_T.
\]
For the second term,
$-u(x_{t+1}-x_t,y_t)=u(x_t,y_t)-u(x_{t+1},y_t)\le d_t$.
Thus $C_T^w\le\Omega_0w_T+\sum_tw_td_t$.  Substituting the rowwise transfer
into the native reduction and dividing the Nash gap by two proves the
theorem. \hfill$\square$

\paragraph{Why the correction is one-sided.}
The learner accumulates $H_\rho$, whereas the target accumulates
$H_\rho-E_\rho$.  For a two-action row with $H_\rho=(r,0)$, positive error
$E_\rho=(e,0)$ with $e\ge0$ cannot increase the target maximum; negative
error $E_\rho=(-e,0)$ can increase it by $e$.  Thus only cumulative
underestimation can create additional positive target regret, exactly as
measured by $\psi(E_\rho)$.  The four numerator terms in
Equation~\ref{eq:weighted-target-debit} have separate roles: delivered learner
regret, public target debit, the endpoint cost of the weighted alternating
observer, and payoff decrease during player-zero updates.  A symmetric norm
is valid but discards the first distinction.  The appendix gives
component-resolved identities that compute or upper-bound the same debit from
retained traces.

\begin{proposition}[Complete coverage is a zero-debit action]
\label{prop:zero-debit-action}
For additive signed RM or RM+ with the uniform zero-positive-mass fallback,
one event of complete, same-origin public
coverage under the event-faithful simultaneous-update contract satisfies
$\epsilon_{\rho,t}=0$ for every row and $d_t=0$.
\end{proposition}

Complete coverage has $B=N$, so its multiplier is one and
$h_{\rho,t}=g_{\rho,t}$ at the same immutable origin; hence
$\epsilon_{\rho,t}=0$.  For the payoff term, consider one updated row and put
$p=\operatorname{RM}(z)$, $h=v-\langle p,v\rangle\one$,
$x=[z]_+$, and $y=[z+h]_+$.  Coordinatewise monotonicity gives
$(y_a-x_a)h_a\ge0$.  If $x$ has positive mass, it is a positive multiple of
$p$; otherwise $x=0$.  In either case $x^Th=0$, hence $y^Th\ge0$.
If $y$ has positive mass, normalization gives
$\langle\operatorname{RM}(z+h),h\rangle\ge0$.  If $y$ has zero mass, the old
mass must also be zero (a positive-mass centered state cannot be projected to
zero), so both policies use the same uniform fallback and the inner product
is zero.  Since $h$ is centered at $p$, this proves
$\langle p^+-p,v\rangle\ge0$ in every mass branch.  RM+ has the same next
policy because applying positive part once more is policy-idempotent.

Let $\sigma$ be the immutable old profile and $\sigma^+$ the simultaneous
update of all rows owned by player zero; the opponent rows agree.  Perfect
recall gives the exact identity
\[
u(\sigma^+,y_t)-u(\sigma,y_t)
=\sum_{I\in\mathcal I_0}\pi_0^{\sigma^+}(I)
 \sum_a(\sigma^+(I,a)-\sigma(I,a))v^\sigma(I,a).
\]
It follows by telescoping terminal-path probabilities and grouping each
changed decision occurrence by its information set; perfect recall makes the
new-policy own reach $\pi_0^{\sigma^+}(I)$ independent of the representative
history.  Every reach factor is nonnegative and every row inner product is
nonnegative by the preceding argument.  Therefore
$u(x_{t+1},y_t)\ge u(x_t,y_t)$, so $d_t=0$.

\subsection{A debit-budgeted controller}

Let $T_k=km$ be aligned checkpoints.  For the executed plan in block $k$,
define its realized weighted charge
\begin{equation}
q_k=
\sum_\rho\psi\!\left(\sum_{t=T_{k-1}+1}^{T_k}
w_t\epsilon_{\rho,t}\right)
+\sum_{t=T_{k-1}+1}^{T_k}w_td_t.
\label{eq:block-charge}
\end{equation}
A partial block has a history-measurable finite envelope $\bar q_k$ satisfying
$0\le q_k\le\bar q_k$; its realized charge may be revealed after execution.
The envelope enters the guarantee and need not be evaluated by the planner.
Complete coverage has $q_k=0$ by
Proposition~\ref{prop:zero-debit-action} and settles immediately.

Let $A_{k-1}$ be the sum of settled partial charges and let $P_{k-1}$ contain
the completed partial blocks whose charges remain unsettled.  Thus the
mathematical realized account is
$Q_{k-1}=A_{k-1}+\sum_{j\in P_{k-1}}q_j$ even when the controller has not yet
observed every term.  Fix a predetermined nonnegative nondecreasing budget
$b(T_k)$ and an outstanding-audit cap $\ell\ge1$.  DB-CFR may select a partial
plan only if
\begin{equation}
A_{k-1}\le b(T_k),\qquad |P_{k-1}|<\ell;
\label{eq:audited-admission}
\end{equation}
otherwise it executes complete coverage.  Within the solvent branch, any
fixed completed-history rule may choose width and order.  Let
$R_k^{(\ell)}$ be the sum of the $\ell$ largest envelopes among partial blocks
executed through $k$, padding with zeros.

\begin{theorem}[Causal exact-fallback guarantee]
\label{thm:debit-controller}
Assume event-faithful candidate executions; a designated complete-coverage
candidate executable from every reachable block boundary; a finite
history-measurable envelope for every selectable partial candidate; globally
zero-initialized additive RM+; centered delivered rows bounded uniformly over
the finite candidate registry; and a fixed nondecreasing output clock.  For
every legal audit settlement order, DB-CFR is well defined,
$Q_K\le b(T_K)+R_K^{(\ell)}$, and
\begin{equation}
\boxed{
\operatorname{expl}(\bar x_{T_K}^w,\bar y_{T_K}^w)
\le
\frac{C_Rw_{T_K}\sqrt{T_K}+\Omega_0w_{T_K}+b(T_K)
+R_K^{(\ell)}}
{2W_{T_K}}.}
\label{eq:controller-bound}
\end{equation}
Under uniform weights, the same statement holds for additive signed RM.
\end{theorem}

\paragraph{Proof.}
The executable complete candidate makes the transition total whenever the
partial admission test fails; otherwise the selected partial candidate and
its finite envelope determine the next event-faithful block.  Thus the stated
controller is well defined under the theorem's premises.

Suppose block $k$ is partial.  Equation~\ref{eq:audited-admission} gives
$A_{k-1}\le b(T_k)$.  Its prior pending blocks together with the new block
form a subset of the executed partial blocks of cardinality at most $\ell$.
Since $q_j\le\bar q_j$ for each member,
\[
Q_k=A_{k-1}+\sum_{j\in P_k}q_j
\le b(T_k)+\sum_{j\in P_k}\bar q_j
\le b(T_k)+R_k^{(\ell)}.
\]
If block $k$ is complete, Proposition~\ref{prop:zero-debit-action} gives
$q_k=0$ and the realized account is unchanged.  Monotonicity of $b$ and of
the top-$\ell$ reserve closes this branch of the induction.  Moving a charge
from pending to settled adds it to $A_k$ and removes the same amount from the
pending sum, so arbitrary audit settlement leaves $Q_k$ invariant.

For each row, $\psi(z)=\max_a[-z(a)]_+$ is subadditive.  Therefore
\[
\sum_\rho\psi\!\left(\sum_kE_{\rho,k}\right)
\le\sum_k\sum_\rho\psi(E_{\rho,k}),
\]
and the nonnegative payoff-decrease charges add exactly.  Hence the two debit
terms in Theorem~\ref{thm:weighted-target-debit} are at most $Q_K$.
Finally, the coordinatewise RM+ clipping telescope gives, for each row with
$\lVert h_{\rho,t}\rVert_2\le G_\rho$,
$[\max_a\sum_t w_th_{\rho,t}(a)]_+
\le w_{T_K}G_\rho\sqrt{T_K}$.  Summing the finite row registry and taking
$C_R=\sum_\rho G_\rho$ yields the stated learner term.  Substitute
$Q_K\le b(T_K)+R_K^{(\ell)}$ into
Theorem~\ref{thm:weighted-target-debit}.  For uniform weights, the separate
positive-part potential argument for unclipped signed RM supplies the same
$C_R\sqrt{T_K}$ form. \hfill$\square$

For a fixed block length, bounded row errors give a finite constant $C_B$ with
$\bar q_k\le C_B\sum_{t=T_{k-1}+1}^{T_k}w_t$.  If $w_t=t^p$, $p\ge0$, and
$b(T)=\kappa w_T\sqrt T$, then
$R_K^{(\ell)}\le\ell C_Bmw_{T_K}$ and
\[
\operatorname{expl}\le\frac{p+1}{2}\left[
\frac{C_R+\kappa}{\sqrt T}
+\frac{\Omega_0+\ell C_Bm}{T}\right].
\]
Thus fixed $\ell$ contributes only $O(T^{-1})$ and preserves the additive-RM+
rate.  The one-block post-execution controller is the case $\ell=1$.  If a
continuation-valid charge is instead available before selection, requiring
$Q_{k-1}+q_k\le b(T_k)$ recovers the zero-overdraft prospective controller as
a special implementation.

\paragraph{Comparison with symmetric charging.}
For every completed block, one-sided debit is no larger than charging
$\sum_\rho\lVert E_\rho\rVert_\infty$ plus the same payoff term.  The
inclusion is strict: the safe plan in the witness below has one-sided charge
zero and symmetric charge $7/4$.  The appendix gives the formal dominance
statement and proof.

\subsection{Why counts cannot certify the decision}

\begin{proposition}[Equal-metadata debit separation]
\label{prop:metadata-separation}
There is a finite two-player zero-sum perfect-recall game and two four-round
singleton public-cut plans $S^+$ and $S^-$ with identical public counts, two
complete coverage cycles, batch width, symbolic work and task graph, maximum
prefix discrepancy $D_4=1$, and scalar strategy-path length $P_4=4$, but
\begin{equation}
q(S^+)=0,\qquad q(S^-)=\frac{11}{8}.
\label{eq:charge-separation}
\end{equation}
Consequently, any charge based only on these shared summaries and sound for
both plans assigns both at least $11/8$; at budget one it rejects $S^+$ even
though its exact debit is zero.  Likewise, a pathwise strict-budget admission
rule restricted to these summaries must reject both before execution.  To
admit the zero-debit plan, a controller must obtain semantic information
before execution or permit a bounded post-execution correction.
\end{proposition}

The witness starts zero-state signed RM on the two public conditional matrices
\[
M_1=\begin{pmatrix}3&3\\1&1\end{pmatrix},\qquad
M_2=\begin{pmatrix}-1&0\\2&-1\end{pmatrix},
\]
and compares the balanced orders $S^+=(1,2,1,2)$ and
$S^-=(2,1,2,1)$.  Exact replay gives
\begin{center}
\small
\begin{tabular}{@{}lcccc@{}}
\toprule
Plan & player-zero error & player-one error & payoff charge & $q$ \\
\midrule
$S^+$ & $(3/4,3/4)$ & $(0,1)$ & $0$ & $0$ \\
$S^-$ & $(-3/8,5/8)$ & $(-3/4,3/4)$ & $1/4$ & $11/8$ \\
\bottomrule
\end{tabular}
\end{center}
Both error rows of $S^+$ are coordinatewise nonnegative.  Plan $S^-$ instead
incurs row underestimation $3/8+3/4$ and payoff decrease $1/4$.  The appendix
gives the full recurrence and rational verification; its symmetric total
charges are $7/4$ and $13/8$, respectively.  The result establishes a finite
decision separation for the declared summary class.

\section{Debit-Budgeted CFR}
\label{sec:method}

DB-CFR separates execution, settlement, and planning.  The \emph{executor}
produces an event-faithful learner record for one declared width and order.
The \emph{auditor} computes the target debit of an executed partial block.
The \emph{planner} reads only completed learner history and settled charges;
it never evaluates an unselected future profile.  Complete coverage connects
these modules by supplying an always-available zero-debit action.

\begin{table}[h]
\caption{One causal DB-CFR block.  Partial plans execute only while the settled
account is solvent and the audit queue has capacity.}
\label{tab:db-cfr-algorithm}
\centering
\small
\begin{tabularx}{\linewidth}{@{}p{0.055\linewidth}X@{}}
\toprule
\multicolumn{2}{@{}l@{}}{\textbf{Input:} learner state, settled account
$A_{k-1}$, pending set $P_{k-1}$, budget $b(T_k)$, cap $\ell$.} \\
\midrule
1 & Settle every completed audit, updating $A_{k-1}$ and $P_{k-1}$. \\
2 & If $A_{k-1}>b(T_k)$ or $|P_{k-1}|\ge\ell$, choose complete coverage. \\
3 & Otherwise choose one registered plan by a fixed completed-history rule. \\
4 & Execute that plan once under the event-faithful contract. \\
5 & Settle a complete block at charge zero; enqueue a partial block for
target-debit audit. \\
\bottomrule
\end{tabularx}
\end{table}

The method is causal because Step 3 cannot read the current block's charge.
Theorem~\ref{thm:debit-controller} permits any such history rule: cost,
previous charges, and learner state may affect the plan, while the debit
account controls when exact recovery is required.  A final reported
certificate either drains the queue or retains the reserve
$R_K^{(\ell)}$.

\subsection{Computing the settled charge}

The charge is defined by the dense target at the learner origins that actually
occurred.  An exact audit evaluates the omitted components at these immutable
origins and applies Equation~\ref{eq:block-charge}; interval or factored
observers may instead upper-bound the same terms.  The appendix gives a finite
summation-by-parts identity that separates the weighted endpoint from component
motion along the executed path.  Complete coverage uses the stronger eventwise
identity $h_{\rho,t}=g_{\rho,t}$ and therefore settles at zero without an
audit.

\subsection{Lagged causal instantiation}

The public-Leduc implementation uses synchronous post-block settlement
($\ell=1$), widths $2$, $5$, and $10$, twelve ten-round blocks, and
$b(T)=12\sqrt T$.  When the account is insolvent it executes width 10.  While
solvent, it observes widths 5 and 2 once, then chooses the least-work partial
width whose most recently settled charge fits the current slack; if neither
does, it chooses the width with the smaller last charge.  The selected block
executes once and its exact charge settles before the next decision.

This rule receives neither exploitability nor a future candidate state.  It
uses zero unselected candidate replays, and its selected width can change with
the realized learner path.  The appendix retains a prospective exact-lookahead
controller as a strong comparator and audits its cost-ordered offer search.

\subsection{Bounded asynchronous settlement}

When selected-path audits run concurrently, the learner may open up to
$\ell$ partial blocks before their charges settle.  Each audit consumes an
immutable event record; the planner enters complete recovery when either the
settled account or the queue cap blocks another partial action.  The finite
reserve in Theorem~\ref{thm:debit-controller} accounts for every hidden charge,
so asynchronous completion changes the bound but not the declared learner
recurrence.  This interface removes audit latency from the sequential decision
path; audit CPU and record traffic remain part of the system cost measured in
Section~\ref{sec:experiments}.

\section{Experiments}
\label{sec:experiments}

The experiments address four questions: whether width and order create a
material target-validity problem, whether target debit yields a nontrivial
control frontier, whether a causal completed-history rule transfers to unseen
public orders, and what its exact audit costs.  They additionally compare
one-sided debit with symmetric error charging.  No controller decision
observes exploitability.

\paragraph{Domains and measures.}
The primary evidence uses released full-range Libratus Turn Subgames 2 and 1
\citep{brown2017libratus}.  Subgame 2 contains 1.86M information sets, 4.78M
actions, 15 public cuts, and 48 river outcomes per cut; Subgame 1 has
$5.7\times$ as many information sets.  Exact exploitability is reported in
milli-big-blinds per game.  The independent domain is ten-card public Leduc,
with a separate Python implementation and ten semantic public-card
components.  Its main quality measure is NashConv, twice
Equation~\ref{eq:gap-definition}.  HUNL comparisons use selected outcomes,
learner CPU under the registered timing protocol, and exact quality.  The
controller comparison reports selected public components in the installed
learner recurrence; post-block audit work is accounted separately.

\paragraph{Non-poker transfer domains.}
Three root-chance-cut games transfer the debit interface out of poker: four-card
imperfect-information Goofspiel (prize-order cut, 24 labels, 680 information
sets), one-die Liar's Dice (joint-deal cut, 36 labels, 24{,}576 information
sets), and 2$\times$2 random-placement Battleship (joint-placement cut, 16
labels, 5{,}588 information sets).  Every terminal belongs to exactly one cut
outcome, so utilities decompose exactly over labels; the campaigns share one
parameterized engine and reproduce the public-Leduc register shapes
(pre-registered seeds, widths dividing the label count, 1e-9 certificate
tolerances, timing inadmissible).  The three-domain matrix is reported by
campaign family only after every domain in that family terminates; null,
negative, and invalid outcomes remain visible under the dual-paper fairness
protocol.  Because non-benign float values trigger the
zero-regret regret-matching discontinuity, the full-coverage dense-reference
parity is pre-registered per domain (learner-state fingerprints at 9 decimals
plus quality agreement, or a documented diagnostic mode); the certificate
identities and full-coverage zero-debit gates are unchanged.

The census and development families are now closed
(Table~\ref{tab:nonpoker-debit}).  Every census is valid and finds both a
nontrivial work--debit frontier and state-dependent partial debit, but no
semantic feature improves over the constant width ranking: the latter is a
three-domain negative result.  Development promotes no controller in
Goofspiel or Battleship.  In Liar's Dice, $\kappa=4$ is the sole promoted
value, winning on three of four development seeds with median matched-work
improvement $0.00626$; thus only Liar's Dice was eligible for a prospective
controller register.
The original 64-epoch target-transfer run proved unsuitable for local CPU
execution and was stopped before any width-2 path completed.  A documented
compute amendment, chosen from Liar's Dice runtime metadata without reading
the six partial Liar's Dice values, replaces it uniformly in all three domains
by four new seeds and coverage epochs $\{1,2,4,8\}$.  The earlier Goofspiel V1
summary had already been inspected, so Goofspiel also receives a new seed
bank.  A restart-semantics audit after the V2 Goofspiel summary then found that
the V2 machine-readable prediction seal retained obsolete V1 absolute counts
despite the registered proportional gates.  V2 is therefore excluded in full.
V3 fixes only that serialization,
uses a shared gate-count function plus fresh seeds, and adds authenticated
per-invocation resume.  All corrected cohorts are valid
(Table~\ref{tab:nonpoker-target}).  Persistent order wins debit on $8/8$
pairs in every domain, but quality on only $1/8$, $4/8$, and $5/8$ pairs in
Goofspiel, Liar's Dice, and Battleship.  Hence every domain supports the
transfer certificates while failing the registered joint ordering
consequence.  The eligible Liar's Dice prospective register required three
fail-closed compute revisions: an oversized V1 runtime pilot, a V2
adjudicator-name error with no terminal result, and a valid one-block V3 pilot
whose horizon could not exercise the development-observed width switch.  The
final V4 register uses the minimum three-block dynamic horizon and six fresh
seeds.  It is valid, pools widths 6 and 36, and has positive median
matched-work improvement $0.01054$, but wins only $4/6$ seeds against the
best fixed control, below the frozen $5/6$ gate.  Prospective transfer is
therefore not established.

\begin{table}[H]
  \caption{Complete non-poker causal-debit census and controller-development
  cohorts.  V, F, and S denote validity, a nontrivial frontier, and
  state-dependent partial debit; R denotes semantic ranking improvement.
  Failed gates are retained rather than filtered.}
  \label{tab:nonpoker-debit}
  \centering
  \footnotesize
  \setlength{\tabcolsep}{4pt}
  \begin{tabular}{@{}lccccc@{}}
    \toprule
    Domain & V & F & S & R & Development decision \\
    \midrule
    Goofspiel-4 & pass & pass & pass & fail & no $\kappa$ promoted \\
    Liar's Dice & pass & pass & pass & fail & $\kappa=4$; $3/4$ wins \\
    Battleship & pass & pass & pass & fail & no $\kappa$ promoted \\
    \bottomrule
  \end{tabular}
\end{table}

\begin{table}[H]
  \caption{Complete corrected non-poker target-transfer and eligible
  prospective cohorts.  Q/D are persistent-order wins for quality/debit out
  of eight paired target paths.  Failed gates are retained.}
  \label{tab:nonpoker-target}
  \centering
  \footnotesize
  \setlength{\tabcolsep}{4pt}
  \begin{tabular}{@{}lccc@{}}
    \toprule
    Domain & Validity & Target Q/D & Prospective decision \\
    \midrule
    Goofspiel-4 & pass & $1/8$ / $8/8$ & unopened (development fail) \\
    Liar's Dice & pass & $4/8$ / $8/8$ & fail: $4/6$ wins; median $+0.01054$ \\
    Battleship & pass & $5/8$ / $8/8$ & unopened (development fail) \\
    \bottomrule
  \end{tabular}
\end{table}

\begin{figure}[t]
\centering
\begin{tikzpicture}
\begin{loglogaxis}[
  name=hunl,
  width=0.40\linewidth,
  height=0.28\linewidth,
  xlabel={exact-equivalent rounds},
  ylabel={exploitability (mbb/g)},
  xlabel near ticks,
  ylabel near ticks,
  tick label style={font=\scriptsize},
  label style={font=\scriptsize},
  scale only axis,
  title style={font=\footnotesize},
  title={(a) HUNL depth crossover},
  legend style={font=\scriptsize, at={(0.98,0.98)}, anchor=north east,
    draw=none, fill=white, fill opacity=0.88, text opacity=1, row sep=-2pt},
  xmin=28, xmax=590,
  ymin=20, ymax=2600,
  log basis x=2,
  xtick={32,64,128,256,512},
  xticklabels={32,64,128,256,512},
  grid=major,
  grid style={gray!18},
]
\addplot+[mark=*, mark size=1.3pt, thick, black]
  coordinates {(32,1814.1) (64,575.5) (128,220.4) (256,84.4) (512,31.1)};
\addlegendentry{exact 48}
\addplot+[mark=square*, mark size=1.2pt, blue!70!black]
  coordinates {(32,1350.5) (64,606.9) (128,345.4) (256,237.6) (512,171.0)};
\addlegendentry{width 24}
\addplot+[mark=triangle*, mark size=1.4pt, red!70!black]
  coordinates {(32,1446.5) (64,719.8) (128,440.2) (256,320.2) (512,233.4)};
\addlegendentry{width 16}
\addplot+[mark=diamond*, mark size=1.4pt, green!45!black]
  coordinates {(32,1479.0) (64,839.3) (128,554.7) (256,388.3) (512,272.7)};
\addlegendentry{width 8}
\draw[densely dashed, gray] (axis cs:48,20) -- (axis cs:48,2600);
\end{loglogaxis}

\begin{axis}[
  at={(hunl.east)}, anchor=west, xshift=8mm,
  width=0.40\linewidth,
  height=0.28\linewidth,
  xlabel={selected components per block},
  ylabel={exact target debit},
  xlabel near ticks,
  ylabel near ticks,
  tick label style={font=\scriptsize},
  yticklabel style={/pgf/number format/fixed,/pgf/number format/precision=0},
  label style={font=\scriptsize},
  scale only axis,
  title style={font=\footnotesize},
  title={(b) public-Leduc debit frontier},
  xmin=25, xmax=215,
  ymin=-2, ymax=57,
  xtick={40,100,200},
  ytick={0,10,20,30,40,50},
  grid=major,
  grid style={gray!18},
]
\draw[blue!70!black, thick] (axis cs:40,9.96) -- (axis cs:40,53.44);
\draw[blue!70!black, thick] (axis cs:100,2.73) -- (axis cs:100,12.33);
\addplot+[only marks, mark=*, mark size=2.0pt, blue!70!black]
  coordinates {(40,22.15) (100,7.18)};
\addplot+[only marks, mark=square*, mark size=2.0pt, black]
  coordinates {(200,0)};
\node[font=\scriptsize, anchor=south west] at (axis cs:41,22.15) {$B=2$};
\node[font=\scriptsize, anchor=south west] at (axis cs:101,7.18) {$B=5$};
\node[font=\scriptsize, anchor=south east] at (axis cs:198,0) {$B=10$};
\end{axis}

\begin{axis}[
  at={(hunl.south west)}, anchor=north west, yshift=-25mm,
  width=0.88\linewidth,
  height=0.22\linewidth,
  ybar,
  bar width=10pt,
  symbolic x coords={1,2,3,4,5,6,7,8,9,10},
  xtick={1,2,3,4,5,6,7,8,9,10},
  xlabel={held-out public-order index},
  ylabel={paired NashConv gain},
  xlabel near ticks,
  ylabel near ticks,
  tick label style={font=\scriptsize},
  yticklabel style={/pgf/number format/fixed,/pgf/number format/precision=2},
  label style={font=\scriptsize},
  scale only axis,
  title style={font=\footnotesize},
  title={(c) lagged causal DB-CFR on unseen orders},
  ymin=-0.01, ymax=0.10,
  ytick={0,0.02,0.04,0.06,0.08,0.10},
  grid=major,
  grid style={gray!18},
]
\addplot+[draw=blue!70!black, fill=blue!55, bar shift=0pt]
  coordinates {(1,0.04527) (2,0.06518) (3,0.06609) (4,0.02301)
    (5,0.03336) (6,0.04288) (7,0.02702) (8,0.04584)
    (9,0.08602) (10,0.08460)};
\draw[black] (axis cs:1,0) -- (axis cs:10,0);
\draw[densely dashed, blue!70!black]
  (axis cs:1,0.04556) -- (axis cs:10,0.04556)
  node[pos=0.50, above=1pt, fill=white, inner sep=1pt,
    font=\scriptsize] {median};
\end{axis}
\end{tikzpicture}
\caption{Existing evidence for the width-control problem and DB-CFR.
\textbf{(a)} Partial HUNL widths lead at 32 exact-equivalent rounds, but
complete coverage overtakes all of them by 64 rounds.  \textbf{(b)} Widths 2,
5, and 10 form a public-Leduc work--debit frontier; points are medians and
vertical segments are ranges.  \textbf{(c)} The prespecified lagged rule improves on
the best no-greater-recurrence-work fixed width on all ten unseen orders.  Gain
is fixed-width NashConv minus DB-CFR NashConv; the dashed line is the median.}
\label{fig:db-cfr-evidence}
\end{figure}

\subsection{Width and order define a real learning frontier}

At 3,072 selected HUNL outcomes per lane, IID singleton sampling reaches
5,042 mbb/g, fresh reshuffling each epoch reaches 2,278 mbb/g, and one
persistent order reaches 1,964 mbb/g.  The persistent improvement over
reshuffling is 314 mbb/g (95\% interval $[282,349]$), although both schedules
cover every outcome once per epoch.  A read-only equal-coverage averaging
control retains 84.7\% of the IID--reshuffle contrast, locating the principal
effect in the delivered feedback path.

With additive RM+, all 21 width--horizon comparisons through 32 exact-
equivalent rounds favor partial coverage on each HUNL input, for 42/42 positive
comparisons.  On Subgame 2 at 1,536 selected outcomes, widths 16 and 24 improve
on complete coverage by 23.3\% and 23.8\%.  A geometric depth ladder then
finds that complete coverage overtakes every partial width between 32 and 64
rounds.  At 512 rounds, complete coverage reaches 31.1 mbb/g, compared with
273, 233, and 171 mbb/g for widths 8, 16, and 24.  The changed ranking across
depth rules out treating one width as a universal estimator setting
(Figure~\ref{fig:db-cfr-evidence}a).

At three exact HUNL quality targets, target-valid IID sampling requires
$1.92$--$2.01\times$ the learner CPU of persistent width 16; exact-to-
persistent ratios are $1.004$--$1.055$.  The appendix reports confidence
intervals, complete grids, the CPU contract, and retained negative controls.

\subsection{The debit census exposes three feasible regimes}

The causal public-Leduc census follows natural alternating signed-RM paths.
Four development orders and three reference widths produce 48 checkpoint
states.  From each state, widths 2, 5, and 10 are replayed for one common
ten-round block, giving 144 paired candidate records.  All centered-row,
nonmutation, deterministic-replay, and dense-parity checks pass.

\begin{table}[h]
\caption{Public-Leduc debit census and causal DB-CFR result.  Work counts
selected public components in the installed recurrence.}
\label{tab:debit-controller}
\centering
\small
\begin{tabular}{@{}lccc@{}}
\toprule
Candidate width & Block work & Debit range & Median debit \\
\midrule
2 & 40 & $9.96$--$53.44$ & $22.15$ \\
5 & 100 & $2.73$--$12.33$ & $7.18$ \\
10 & 200 & $0$ & $0$ \\
\midrule
\multicolumn{4}{@{}l@{}}{Lagged causal DB-CFR: 10/10 unseen-order wins;
median NashConv gain $0.0456$; median work 720.}\\
\bottomrule
\end{tabular}
\end{table}

All three widths appear on the exact work--debit frontier
(Table~\ref{tab:debit-controller}).  Width 5 has lower debit than width 2 in
all 48 contexts, while debit magnitude changes markedly with learner state.
Thus the controller decision is not which partial width has smaller debit; it
is which frontier point fits the remaining cumulative budget.

\paragraph{Symmetric-charge comparison.}
The symmetric comparator is strictly larger on all 96 partial records (median
ratio $2.325\times$).  Under the ten-round budget induced by $\kappa=12$,
one-sided debit admits 84 partial offers versus 56; both charges remain zero
for complete coverage.

\subsection{A causal controller transfers to unseen orders}

The causal study fixes $b(T)=12\sqrt T$ and twelve ten-round blocks.  On four
development orders, five completed-history policies produce a work--quality
Pareto set containing fixed width 5, the lagged rule from
Section~\ref{sec:method}, and exact coverage.  The lagged rule is prespecified
from development; it changes width while using substantially less
recurrence work than fixed width 5.  No policy evaluates an unselected
continuation.

On ten disjoint held-out public orders, lagged DB-CFR beats the lowest-NashConv
fixed width using no more installed component work on all ten pairs
(two-sided exact sign test $p=0.00195$; Figure~\ref{fig:db-cfr-evidence}c).
The paired median gain is 0.0456.  The
controller uses widths 2, 5, and 10, produces ten distinct schedules, and uses
660--780 components (median 720).  All finite-overdraft, exact-fallback,
matched-work, and fingerprint checks pass.  Appendix
Table~\ref{tab:app-causal-controller} reports the complete register.

The result changes the learner path without future candidate replay.  It
isolates debit-guided recurrence work; the full-target post-block audit and a
prospective exact-lookahead comparator are reported separately in the appendix.

\paragraph{Audit resources.}
Across nine preregistered HUNL width--seed cells, the exact target-debit core's
incremental process CPU equals a median 1.56 and maximum 3.43 matched
learner-only runs, meeting the registered periodic-audit thresholds.  Omitted-
component quotient traversals account for 88.2\% of that cost.

\section{Related Work}
\label{sec:related}

MCCFR, Public Chance Sampling, and stochastic CFR obtain target regret from
conditionally unbiased feedback
\citep{lanctot2009mccfr,johanson2012pcs,farina2020stochastic}; DB-CFR instead
controls deterministic history-coupled bias by pathwise target debit.  Lazy CFR
skips information-set updates, while GraB and random reshuffling order smooth
finite-sum gradients \citep{zhou2020lazy,lu2022grab,das2022withoutreplacement}.
Delayed-feedback learning postpones the learner update
\citep{joulani2013delay}; here only the target-validity audit is delayed.
Persistent low-discrepancy MCCFR studies learner-specific thinning of sparse
node streams in a concurrent submission \citep{anonymous2027amccfr}, not width
over an explicit public cut.  Appendix~\ref{app:related-work} compares the
interfaces in detail.

\section{Limitations and Conclusion}
\label{sec:limitations}

\paragraph{Conclusion.}
Conditional unbiasedness is sufficient but not necessary for target-valid CFR
feedback at an explicit finite public cut.  One-sided target debit supplies a
pathwise alternative: it transfers delivered regret to the dense target game,
makes complete coverage a zero-debit recovery action, and supports causal
width and order decisions without future candidate replay.  The equal-metadata
converse identifies the required semantic information, while the two-domain
evidence connects it to a width/order crossover and a held-out dynamic-width
consequence.  Public-feedback granularity can therefore be treated as a
controlled learning decision with an explicit target-game account.

\paragraph{Scope and limitations.}
The theory covers finite additive public cuts, event-faithful alternating
signed RM under uniform weights, and RM+ under a predetermined nondecreasing
output clock.  Dynamic control is evaluated in public Leduc; HUNL establishes
the frontier but not an end-to-end DB-CFR speedup.  The matched-work result
counts installed recurrence components and reports its exact selected-path
audit separately; a full-state asynchronous prototype is limited by snapshot
and memory traffic.  Nested cuts and other learner recurrences require separate
transfer arguments.  Compact within-origin factorization is the next step
toward online DB-CFR in large HUNL endgames.

\bibliography{references}
\bibliographystyle{iclr2027_conference}

\appendix
\section{Extended Related Work}
\label{app:related-work}

Table~\ref{tab:main-separation} compares the closest theorem and method
interfaces.  The relevant distinction is the complete decision tuple, not the
shared vocabulary of sampling, batching, or component order.

\begin{table}[h]
\caption{Closest interfaces to debit-budgeted public feedback.}
\label{tab:main-separation}
\centering\footnotesize
\begin{tabularx}{\linewidth}{@{}p{0.16\linewidth}XX@{}}
\toprule
Interface & Established conclusion & Missing interface component \\
\midrule
MCCFR and Public Chance Sampling
& Regret from randomly sampled terminal-history or public-chance blocks.
& Their guarantees use conditional target validity; they do not control a
deterministic history-coupled public error or select partial versus exact
coverage. \\
\addlinespace
Generalized and stochastic CFR
& Modular regret minimization with bounded unbiased value or gradient
estimators.
& The martingale comparison fails when a without-replacement public order is
coupled to the evolving learner state. \\
\addlinespace
Lazy CFR
& Reach-aware skipping and aggregation of information-set updates.
& It changes which information sets update, not which public components form
one target row; it has no dense target-debit budget. \\
\addlinespace
GraB and random reshuffling
& History-dependent component order and discrepancy-based rates in smooth
finite-sum optimization.
& They do not supply counterfactual local-regret reduction, the alternating
output bridge, or a zero-debit public action. \\
\addlinespace
Aligned persistent MCCFR
& Guarantees on owner-thinned sparse chance streams in External Sampling.
& It addresses a persistent-state key under opponent sampling, not width and
order over one explicit finite public cut. \\
\bottomrule
\end{tabularx}
\end{table}

\paragraph{Sampling-based CFR.}
MCCFR partitions terminal histories and constructs a sampled counterfactual
value whose expectation equals the dense value
\citep{lanctot2009mccfr}.  Public Chance Sampling specializes the partition to
public outcomes and exploits vectorized reuse over private chance
\citep{johanson2012pcs}.  Generalized sampling permits any bounded unbiased
value estimator and makes estimator variance explicit
\citep{gibson2012generalized}; stochastic regret minimization similarly
combines a regret minimizer with conditionally unbiased gradient estimates
\citep{farina2020stochastic}.  A deterministic remaining batch in a
without-replacement epoch has no such eventwise conditional law.  DB-CFR
instead compares the realized delivered sum directly with the dense target
sum and charges only target-regret underestimation.

\paragraph{Order and delayed computation.}
Random reshuffling, quasi--Monte Carlo integration, and GraB connect component
order or discrepancy with temporal variation
\citep{niederreiter1992qmc,lu2022grab,das2022withoutreplacement}.  Lazy CFR
uses reach structure to update a subset of information sets while aggregating
their skipped counterfactual rewards \citep{zhou2020lazy}.  These are important
ingredients and analogies.  The nonroutine step here is the full-game
composition: centered public components are evaluated at native alternating
origins, transferred through local counterfactual regret, and constrained by
a causal one-sided debit with an exact feasible action.

Online learning with delayed feedback instead postpones the loss, gradient,
or reward needed for the learner update
\citep{joulani2013delay,flaspohler2021delay}.  In DB-CFR, the selected partial
row reaches RM/RM+ immediately and changes the next policy; only the separate
dense-target audit is unsettled.  The outstanding correction therefore enters
the delivered-to-target transfer as a debit reserve rather than as delayed
learner regret.  This distinction also separates the theorem from generic
asynchronous CFR implementations, which change when regret updates are
applied.

\paragraph{Approximate feedback and execution.}
Functional and approximate regret matching charge per-round input error
\citep{waugh2015functional,dorazio2019approximate}.  Such a symmetric charge
is valid but can discard the signed cancellation within a balanced public
epoch.  Equation~\ref{eq:app-weighted-deadline-identity} retains this
cancellation.
Dataflow systems and parallel CFR supply grouping, progress, and physical
execution mechanisms \citep{akidau2015dataflow,murray2013naiad,kim2026parallel};
the event-faithful contract uses those ideas only to preserve one declared
learner transition.  They do not define the target-game row or its regret
transfer.

\paragraph{Concurrent persistent MCCFR.}
A concurrent anonymous submission studies a distinct failure mode:
alternating External-Sampling traversals adaptively thin a persistent
concrete-node chance stream into learner-specific subsequences
\citep{anonymous2027amccfr}.  Its owner--node key, sampled-tree recursion, and
opponent martingale do not appear here.  Conversely, DB-CFR's public width,
one-sided debit, complete-coverage fallback, HUNL frontier, and public-Leduc
controller do not appear in that work.  Both use finite summation by parts as
classical algebra.

\section{Proof Details}
\label{app:proofs}

\paragraph{Proof dependency map.}
Lemma~\ref{lem:app-public-cut-lanes} establishes the additive component
identity used throughout.  The weighted alternating reduction and one-sided
row inequality prove Theorem~\ref{thm:weighted-target-debit}; blockwise
subadditivity, exact payoff monotonicity, and the bounded-settlement induction
prove
Theorem~\ref{thm:debit-controller}.  Proposition
\ref{prop:symmetric-charge-separation} follows from the pointwise comparison
between the negative-part gauge and the infinity norm.  Proposition
\ref{prop:app-public-order-witness} supplies the equal-metadata witness used by
Proposition~\ref{prop:metadata-separation}.  The local component identity leads
to Theorem~\ref{thm:app-weighted-component-certificate} and the reported-output
certificate in Equation~\ref{eq:app-complete-weighted-output-certificate}.
Lemma~\ref{lem:app-rm-log-path} supplies the execution-dependent path rate used
only by the appendix structural corollaries.

\paragraph{Mechanized scope.}
The accompanying Lean development under \texttt{formal/target\_transfer}
checks the finite-EFG semantics, perfect-recall local-to-external reduction,
native alternating-origin correction, bilinear realization-plan average,
weighted additive-RM+ envelope, exact complete-coverage monotonicity, and the
top-$\ell$ controller account. Its composed exact-feedback theorem has the
same $2W_T$ normalization used below. The mechanization averages realization
vectors, not behavioral rows componentwise. It also checks the exact rowwise
delivered-to-target debit composition, the finite vector Abel identity, its
endpoint--variation norm bound, and the factor-two weighted-prefix estimate.
The uniform-weight signed-RM envelope is checked separately with the squared
positive-part potential of its unclipped signed state, rather than being
inferred from RM+. The standard local own-reach numerator/denominator average
is proved to reconstruct the normalized realization-vector sum on consistent
own sequences, including the zero-prefix uniform fallback; concrete tree-map
and floating-point refinement remain outside the current certificate.
Concrete Python/C++ accumulator and planner/executor refinement, and the
solver-specific public-component incidence contract, remain explicit artifact
boundaries: the generic Abel results do not by themselves prove that the
concrete cut executor supplies the declared component streams.
For finite root-chance cuts, Lean now also checks that the chance-weighted
forced-root components sum to the dense counterfactual row before and after
centering, and that an explicitly complete label partition preserves this
identity. Matching concrete traversal arrays and scaling to those formal
components remains the program-refinement boundary.
Lemma~\ref{lem:app-rm-log-path} is mechanized through the finite-horizon
statement: Lean checks positive-energy growth and persistence, the normalized
angular bound, finite-dimensional $\ell_1/\ell_2$ conversion, each squared
$4d\log r$ policy toll, the adjacent-log telescope, temporal
Cauchy--Schwarz, the cumulative terminal-energy bound, and the separate
simplex-diameter activation charge.  One positive-part recurrence contract
covers both additive signed RM and RM+; separate Lean corollaries instantiate
that contract with, respectively, the unprojected additive recurrence and the
coordinatewise-clipped RM+ recurrence.

\subsection{Deterministic balanced public-chance CFR}
\label{app:balanced-cfr-proof}

We first fix the clocks used throughout this proof.  Work in real arithmetic
in one finite two-player zero-sum perfect-recall extensive-form game (EFG).
Every physical realization satisfies the event-faithful contract of
Section~\ref{sec:background}; the proof therefore operates on the resulting
logical event sequence.  The native alternating
chain is
\begin{equation}
(x_t,y_t)\longrightarrow(x_{t+1},y_t)
\longrightarrow(x_{t+1},y_{t+1}),
\label{eq:app-alternating-chain}
\end{equation}
so player zero consumes feedback at $(x_t,y_t)$ and player one at
$(x_{t+1},y_t)$.  Each player and public lane has its own player-event clock.
For every owned row $\rho$, require the exact additive identity
\begin{equation}
g_{\rho,t}=g^{\rm rem}_{\rho,t}+\sum_{\ell\in L_\rho}g_{\rho,\ell,t},
\qquad
h_{\rho,t}=g^{\rm rem}_{\rho,t}+\sum_{\ell\in L_\rho}h_{\rho,\ell,t}.
\label{eq:app-additive-lanes}
\end{equation}
Every terminal contribution occurs exactly once.  Lane $\ell$ has
$N_\ell$ labels, width $B_\ell\mid N_\ell$, and
$m_\ell=N_\ell/B_\ell$.  Each lane uses a declared owner-event clock; its
batch sequence may be arbitrary.  Exact consecutive coverage is imposed only
in the balanced corollaries below.  Lane clocks need not align.  Let
$\mathcal S_\rho$ be the finite row/lane streams incident to row $\rho$, and
let $\tau_s(n)$ be stream $s$'s strictly increasing map from local
event $n$ to a global owner round.  The clock contract is the incidence
identity
\begin{equation}
h_{\rho,t}-g_{\rho,t}
=\sum_{s\in\mathcal S_\rho}\sum_{n:\tau_s(n)=t}\sum_j
a_{s,n,j}X_{\rho,s,n,j}.
\label{eq:app-stream-incidence}
\end{equation}
Let $i(s)$ denote the owner of stream $s$.  Every stream incident to row
$\rho=(i,I)$ satisfies $i(s)=i$.  Its components
are evaluated at that row owner's native immutable consuming profile in round
$\tau_s(n)$, and all components incident to one
owner round are aggregated before its single learner update.  If no stream
event is incident, its contribution to the target/delivered \emph{difference}
is exactly zero.  This identity is the bridge between local stream clocks and
the global regret sums.

For a row $\rho=(i,I)$ and lane $\ell$, let
$X_{i,I,\ell,n,j}$ be public label $j$'s target-chance-weighted conditional
counterfactual contribution at the immutable consuming profile.  With
$S_{i,\ell,n}$ denoting the selected batch, define
\begin{align}
a_{i,\ell,n,j}
  &=m_\ell\ind{j\in S_{i,\ell,n}}-1,
&A_{i,\ell,q,j}&=\sum_{n=1}^q a_{i,\ell,n,j},
\label{eq:app-public-coefficients}
\end{align}
At global horizon $T$, let $n_{i,\ell}(T)$ count consumed local events and set
\begin{equation}
D_T=\max_{i,\ell,j,\,0\le q\le n_{i,\ell}(T)}|A_{i,\ell,q,j}|.
\label{eq:app-joint-discrepancy}
\end{equation}
Let $g_{i,I,t}$ be the exact target row and
$h_{i,I,t}$ the delivered scaled row.  Both are centered at the
same consuming behavior; concretely, an uncentered label contribution $v_j$
induces $X_j=v_j-\langle p,v_j\rangle\mathbf 1$.  A physically omitted row may
be the conceptual callback $h_{i,I,t}=0$ only when the declared delivered row
is semantically zero.  For additive RM/RM+ such a zero row leaves learner
state and policy unchanged; other recurrences need a separate semantic audit.

Nested partial cuts generally violate
Equation~\ref{eq:app-additive-lanes}.  If two Horvitz--Thompson (HT) weights
are $w_1,w_2$, then
$(w_1w_2-1)X=(w_1-1)X+(w_2-1)X+(w_1-1)(w_2-1)X$; independent lane ledgers
omit the interaction.  Such cuts require a joint registry or an explicit
interaction tensor.

\begin{lemma}[Nonnested public cuts induce fixed counterfactual lanes]
\label{lem:app-public-cut-lanes}
Let registered public chance nodes have finite label sets $J_\ell$, and let
$\mathcal Z_{\ell,j}$ be the terminals descending from labelled branch
$(\ell,j)$.  Suppose
\begin{equation}
\mathcal Z=\mathcal Z^{\rm rem}\mathbin{\dot\cup}
\mathop{\dot\bigcup}_{\ell,j}\mathcal Z_{\ell,j},
\label{eq:app-public-terminal-partition}
\end{equation}
so every terminal crosses zero or one registered branch.  Fix row
$\rho=(i,I)$ and one immutable native consuming profile $\sigma_{i,t}$.  For
$a\in A(I)$ define the standard unnormalized counterfactual terminal kernel
\begin{equation}
K_{\rho,t,a}(z)=
\sum_{\substack{h\in I\\ha\sqsubseteq z}}
\pi_{-i}^{\sigma_{i,t}}(h)\pi^{\sigma_{i,t}}(ha,z)u_i(z).
\label{eq:app-counterfactual-terminal-kernel}
\end{equation}
Let $V^{\rm rem}_{\rho,t}$ and $V_{\rho,\ell,t,j}$ sum this action vector over
$\mathcal Z^{\rm rem}$ and $\mathcal Z_{\ell,j}$, respectively.  With
$C_p(v)=v-\langle p,v\rangle\one$, put
$g^{\rm rem}_{\rho,t}=C_{p_{\rho,t}}(V^{\rm rem}_{\rho,t})$ and
$X_{\rho,\ell,t,j}=C_{p_{\rho,t}}(V_{\rho,\ell,t,j})$.  Then
\begin{align}
g_{\rho,t}&=g^{\rm rem}_{\rho,t}+\sum_{\ell,j}X_{\rho,\ell,t,j},
\nonumber\\
h_{\rho,t}-g_{\rho,t}
&=\sum_{\ell,j}\left(m_\ell\ind{j\in S_{i,\ell,t}}-1\right)
X_{\rho,\ell,t,j}
\label{eq:app-public-cut-delivery-identity}
\end{align}
whenever a width-$B_\ell$ traversal retains the remainder exactly and scales
selected labels by $m_\ell=|J_\ell|/B_\ell$.  This identity is deterministic;
under uniform sampling the same multiplier is the Horvitz--Thompson weight.
Structurally absent, unreachable, or private-card-blocked components are zero.
\end{lemma}

\paragraph{Proof.}
Standard unnormalized counterfactual action value is the sum of
Equation~\ref{eq:app-counterfactual-terminal-kernel} over terminals.  The
disjoint partition decomposes it into the remainder plus the lanes, and $C_p$
is linear.  A selected traversal changes branch $(\ell,j)$'s coefficient from
one
to $m_\ell\ind{j\in S_{i,\ell,t}}$ and leaves the remainder unchanged;
subtraction proves Equation~\ref{eq:app-public-cut-delivery-identity}.  Chance
reach appears exactly once, in the prefix if the cut precedes $I$ and in the
continuation otherwise.  The two alternating owner events apply the same
argument at their different immutable origins. \hfill$\square$

For native private-hand row $k$, multiplying by its fixed own-root factor and
dividing by compatible joint mass and utility scale commutes with every sum
above; the own-root factor must remain inside the row sum.  Equation
\ref{eq:app-public-cut-delivery-identity} pulled through $\tau_s$ gives
Equation~\ref{eq:app-stream-incidence}.  A conceptual zero callback is
state-identical to omission for additive signed RM/RM+, but not automatically
for discounting or predictive recurrences.

Let $d_1$ sum rowwise policy $\ell_1$ distances over the fixed information-set
registry.  If $\sigma^0,\ldots,\sigma^{2T}$ are the profiles before, between,
and after the two updates in Equation~\ref{eq:app-alternating-chain}, define
\begin{equation}
P_T=\sum_{n=1}^{2T}d_1(\sigma^{n-1},\sigma^n).
\label{eq:app-projective-path}
\end{equation}
Let $P_{i,T}$ denote the contribution from player $i$'s transitions, so
$P_T=P_{0,T}+P_{1,T}$.

\begin{proposition}[Exact balance and conditional unbiasedness are distinct]
\label{app:balance-vs-unbiasedness}
Let $J$ contain $N$ labels, let $B\mid N$, and put $m=N/B$.  At event $t$,
let $S_t\subseteq J$ have size $B$, and let $\mathcal F_{t-1}$ contain the
completed learner history and previous selections.  If the bounded immutable
component array $X_{t,j}$ is $\mathcal F_{t-1}$-measurable, define
\[
H_t=m\sum_{j\in S_t}X_{t,j},\qquad
G_t=\sum_{j\in J}X_{t,j},\qquad
\pi_{t,j}=\Pr(j\in S_t\mid\mathcal F_{t-1}).
\]
Then
\begin{equation}
\mathbb E[H_t-G_t\mid\mathcal F_{t-1}]
=\sum_{j\in J}(m\pi_{t,j}-1)X_{t,j}.
\label{eq:app-balance-unbiased}
\end{equation}
Thus conditional unbiasedness uniformly over all bounded current arrays holds
iff $\pi_{t,j}=1/m$ for every label.  Exact epoch coverage and this eventwise
condition imply neither one another.
\end{proposition}

\paragraph{Proof.}
Equation~\ref{eq:app-balance-unbiased} is conditional linearity.  Necessity
follows by choosing a scalar array supported on one label, and sufficiency is
immediate.  For balance without eventwise unbiasedness, take $N=2,B=1$ and
either exactly covered order.  After the first selection the remaining label
has conditional inclusion probability one, not $1/2$.  Conversely, independent
uniform $B$-subsets at every event have inclusion probability $B/N=1/m$, but
repeat or omit labels over an $m$-event block with positive probability.
\hfill$\square$

\begin{proposition}[Unique universal zero-debit width]
\label{prop:app-unique-zero-debit-width}
For a fixed-size public batch $S\subseteq J$, the pathwise identity
$m\sum_{j\in S}X_j=\sum_{j\in J}X_j$ holds for every finite component array
$(X_j)_{j\in J}$ if and only if $B=N$ (equivalently $m=1$ and $S=J$).
Thus complete coverage is the unique width whose public debit vanishes at
every event independently of the current CFR profile.
\end{proposition}

\paragraph{Proof.}
Universal equality requires every coefficient
$m\ind{j\in S}-1$ to be zero: choose an array supported on any one label.
An omitted label has coefficient $-1$, so none can be omitted; then $S=J$,
$B=N$, and $m=1$.  The converse is immediate.  \hfill$\square$

\begin{lemma}[Finite RM+ phase capture]
\label{lem:app-rmplus-phase-capture}
Let a two-action additive-RM+ row start a phase at nonnegative state $q$ with
$Q=\norm{q}_2^2$.  Suppose the same action is better throughout the phase by
an advantage in $[a,b]$, where $0<a\le b$.  Its regret-matching probability
for that action is monotone and becomes one within
\begin{equation}
N(Q;a,b)=\left\lceil\frac{4\sqrt Q}{a}\right\rceil
+\left\lceil\frac{4b^2}{a^2}\right\rceil+1
\label{eq:app-rmplus-capture}
\end{equation}
updates.
\end{lemma}

\paragraph{Proof.}
Orient the desired action first.  Until the other score clips, its score
difference increases by the advantage and the desired probability is
monotone.  Centering gives $q^\top h=0$, while projection can only decrease
norm, so after $n$ updates $Q_n\le Q+nb^2$.  If capture has not occurred,
\[
-\sqrt Q+na\le q_{n,0}-q_{n,1}<\sqrt{Q+nb^2}
\le\sqrt Q+b\sqrt n.
\]
For the integer in Equation~\ref{eq:app-rmplus-capture}, Young's inequality
$b\sqrt n/a\le n/2+b^2/(2a^2)$ makes the left bound strictly exceed the
right, a contradiction. \hfill$\square$

\begin{proposition}[RM+ product-scale necessity]
\label{prop:app-discrepancy-path-necessity}
There is a fixed rational finite two-player zero-sum perfect-recall EFG and a
fixed infinite precommitted public order for which both players run
zero-state additive RM+ at their native alternating origins and
\begin{equation}
\widehat{\mathcal R}_T\le6\sqrt T,\qquad
D_T/T\to0,\qquad P_T/T\to0.
\label{eq:app-rmplus-necessity-global}
\end{equation}
Nevertheless, at completed macrocycle horizons $T_K$,
\begin{equation}
D_{T_K},P_{T_K}=\Theta(\sqrt{T_K}),\qquad
\frac{D_{T_K}P_{T_K}}{T_K}\to4,\qquad
\liminf_K\operatorname{expl}(\bar x_{T_K},\bar y_{T_K})\ge\frac{11}{64},
\label{eq:app-rmplus-necessity-endpoints}
\end{equation}
and cumulative delivered-minus-target row error is $\Omega(D_{T_K}P_{T_K})$.
More precisely, fix finite $c_D,c_P\ge0$.  If a remainder $r_T$ makes
\begin{equation}
2T\operatorname{expl}(\bar x_T,\bar y_T)
\le \widehat{\mathcal R}_T+c_DD_T+c_PP_T+r_T
\label{eq:app-rmplus-remainder-form}
\end{equation}
valid for every execution in the declared zero-state additive-RM+ class, then
on this fixed execution
\begin{equation}
\liminf_{K\to\infty}
\frac{r_{T_K}}{D_{T_K}P_{T_K}}\ge\frac{11}{128}.
\label{eq:app-rmplus-remainder-lower}
\end{equation}
Thus a target-transfer correction $o(D_TP_T)$ cannot hold uniformly even for
this RM+ recurrence.  The claim is about low normalized discrepancy with
macrocycle closure, not exact consecutive balance, and does not cover signed
RM.
\end{proposition}

\paragraph{Proof.}
Player zero chooses $A/B$; player one, without observing that action, chooses
$L/R$; terminal public chance selects $j\in\{1,2\}$ uniformly.  Player-zero's
two payoff matrices and their exact target average are
\begin{equation}
U_1=\begin{pmatrix}-4&-2\\-2&0\end{pmatrix},\qquad
U_2=\begin{pmatrix}6&4\\2&0\end{pmatrix},\qquad
\bar U=\frac{U_1+U_2}{2}=\begin{pmatrix}1&1\\0&0\end{pmatrix}.
\label{eq:app-necessity-game}
\end{equation}
Player one receives the negative payoff.  Selecting one of the two labels
uses its chance weight $1/2$ and HT multiplier two, hence delivers the
centered row of $U_j$.  Label one gives player zero $A-B$ advantage $-2$ and
player one $L-R$ advantage $2$; label two gives advantages $4$ and $-2$.
They are independent of the opponent policy.  The same precommitted order is
used on the players' distinct owner clocks, so player zero reads $(x_t,y_t)$
and player one reads $(x_{t+1},y_t)$ exactly as required.

Fix $C=4096$.  Macrocycle $k$ is
\begin{equation}
1^{Ck},\qquad 2^{2Ck},\qquad 1^{Ck}.
\label{eq:app-necessity-order}
\end{equation}
Each label's coefficient prefix returns to zero at the cycle endpoint and
has maximum magnitude $Ck$.  Therefore
\begin{equation}
T_K=2CK(K+1),\qquad D_{T_K}=CK.
\label{eq:app-necessity-clocks}
\end{equation}
For either two-action row, a centered callback with advantage magnitude at
most $b$ increases squared RM+ potential by at most $b^2$.  Consequently
$Q_{0,t}\le16t$ and $Q_{1,t}\le4t$.  Since $t\le T_k\le4Ck^2$ inside cycle
$k$, Lemma~\ref{lem:app-rmplus-phase-capture} bounds the slowest capture by
$16\sqrt{C}\,k+5=1024k+5<Ck$.  Every phase therefore reaches its designated
pure pair: $(B,L)$ on label one and $(A,R)$ on label two.  The initial joint
uniform-to-pure move costs two in rowwise $\ell_1$ path and the two joint
pure switches per cycle cost four each, giving
\begin{equation}
P_{T_K}=8K+2,\qquad
\frac{D_{T_K}P_{T_K}}{T_K}=\frac{8K+2}{2(K+1)}\to4.
\label{eq:app-necessity-path}
\end{equation}
An unfinished cycle has $O(K)$ events and $O(1)$ additional path against an
$\Omega(K^2)$ completed prefix, which proves both normalized limits in
Equation~\ref{eq:app-rmplus-necessity-global} at all horizons.

Reflection and the centered-potential calculation give
$\widehat{\mathcal R}_T\le\norm{q_{0,T}}_2+
\norm{q_{1,T}}_2\le6\sqrt T$.  Let
$B_K=\sum_{t\le T_K}x_t(B)$.  The ideal pure phases contribute $T_K/2$.
Counting the pre-update capture event, total transition contamination obeys
\begin{equation}
\left|B_K-\frac{T_K}{2}\right|
\le20\sqrt C K(K+1)+15K,
\label{eq:app-necessity-contamination}
\end{equation}
because the two label-one phases cost at most $16\sqrt{C}\,k+5$ each and the
label-two phase at most $8\sqrt{C}\,k+5$.  Hence
$\liminf B_K/T_K\ge1/2-10/\sqrt C=11/32$.

The columns of $\bar U$ coincide and $A$ dominates $B$ by one.  Thus the
uniform pre/pre output has
$\operatorname{Gap}=B_K/T_K$ and
$\operatorname{expl}=B_K/(2T_K)$, proving the last part of
Equation~\ref{eq:app-rmplus-necessity-endpoints}.  The target cumulative
player-zero $A$-coordinate equals $B_K$, whereas its delivered coordinate is
at most its positive delivered regret, at most $4\sqrt{T_K}$.  The
delivered-minus-target row norm is therefore at least
$B_K-4\sqrt{T_K}=\Omega(T_K)=\Omega(D_{T_K}P_{T_K})$.
For Equation~\ref{eq:app-rmplus-remainder-form}, divide by $T_K$.  The
delivered-regret and fixed linear-discrepancy/path terms vanish, while
$2\operatorname{expl}$ has liminf at least $11/32$.  Hence
$\liminf_K r_{T_K}/T_K\ge11/32$; combine this with
$D_{T_K}P_{T_K}/T_K\to4$ to obtain
Equation~\ref{eq:app-rmplus-remainder-lower}.  All matrices and the order are
fixed independently of the reporting horizon.
\hfill$\square$

For an exactly covered epoch, $a_{r,j}=m\ind{j\in S_r}-1$ has
$A_{m,j}=0$.  Summation by parts therefore gives the different, pathwise
identity
\begin{equation}
\sum_{r=1}^m(H_r-G_r)
=-\sum_{j\in J}\sum_{r=1}^{m-1}
A_{r,j}(X_{r+1,j}-X_{r,j}).
\label{eq:app-balance-pathwise}
\end{equation}
Balance cancels fixed components exactly and charges evolving components by
variation; it is not a martingale-difference argument.

\begin{lemma}[Discrepancy--path public transfer]
\label{lem:app-balanced-error}
There are finite structural constants $C_0,C_X$, chosen before public order
is realized, such that for every horizon
\begin{equation}
\sum_{i,I}\norm{\sum_{t=1}^T(h_{i,I,t}-g_{i,I,t})}_\infty
\le C_0D_T+C_XD_TP_T.
\label{eq:app-balanced-error}
\end{equation}
The statement is pathwise and requires neither independence nor conditional
unbiasedness.
\end{lemma}

\paragraph{Proof.}
By Equation~\ref{eq:app-stream-incidence}, the global error sum is the sum of
the incident local-stream errors.  For one row and stream, suppress fixed
indices, put $q=n_s(T)$, and write its contribution as
$\sum_{n=1}^{q}\sum_j a_{n,j}X_{n,j}$.  When $q=0$ it is zero.  Otherwise,
discrete summation by parts on the local clock gives
\begin{equation}
\sum_{n=1}^{q}\sum_j a_{n,j}X_{n,j}
=\sum_j A_{q,j}X_{q,j}
-\sum_{n=1}^{q-1}\sum_j A_{n,j}(X_{n+1,j}-X_{n,j}).
\label{eq:app-abel}
\end{equation}
Because the game is finite, choose label-aggregated constants satisfying
\begin{align}
\sup_\sigma\sum_j\norm{X_{\rho,\ell,j}(\sigma)}_\infty
&\le X^{\max}_{\rho,\ell},\\
\sum_j\norm{X_{\rho,\ell,j}(\sigma')-
X_{\rho,\ell,j}(\sigma)}_\infty
&\le L^X_{\rho,\ell}d_1(\sigma',\sigma).
\end{align}
For standard unnormalized finite-EFG counterfactual components, the maps are
multiaffine and have bounded derivatives on the compact behavior polytope.
Take
\begin{align}
C_0&=\sum_{\rho,\ell}X^{\max}_{\rho,\ell},&
C_X&=\sum_{\rho,\ell}L^X_{\rho,\ell}.
\label{eq:app-structural-constants}
\end{align}
The endpoint is at most $C_0D_T$.  Consecutive origins of one stream delimit
disjoint global path segments, so its variation is at most
$D_TL^X_{\rho,\ell}P_T$; finite cross-stream reuse is absorbed by the sum in
$C_X$.  This also covers every stopped horizon. \hfill$\square$

\begin{lemma}[Logarithmic projective path for additive RM/RM+]
\label{lem:app-rm-log-path}
Consider one zero-state $d$-action row with either recurrence
\begin{equation}
z_t=z_{t-1}+h_t
\quad\text{or}\quad
z_t=[z_{t-1}+h_t]_+,
\qquad p_t=\operatorname{RM}(z_{t-1}),\qquad
\langle p_t,h_t\rangle=0,\qquad \norm{h_t}_2\le G.
\label{eq:app-rm-row}
\end{equation}
If the positive potential first becomes nonzero at time $\tau$, write
$q=\norm{[z_\tau]_+}_2^2>0$.  Then, for every $T\ge\tau$,
\begin{equation}
\sum_{t=1}^T
\norm{\operatorname{RM}(z_t)-\operatorname{RM}(z_{t-1})}_1
\le 2+2\sqrt{dT\log\!\left(1+\frac{TG^2}{q}\right)}.
\label{eq:app-rm-path-bound}
\end{equation}
A row that never activates stays uniform and has zero path.  Consequently, a
fixed finite additive-RM or RM+ game satisfies
$P_T=O_e(\sqrt{T\log(T+1)})$ on every fixed infinite execution, where $O_e$
permits its constant to depend on the execution's first positive row
potentials.
\end{lemma}

\paragraph{Proof.}
Set $u=[z_{t-1}]_+$, $v=[z_t]_+$,
$Q_{t-1}=\norm{u}_2^2$, and $Q_t=\norm{v}_2^2$.  On every coordinate where
$u_a>0$, both recurrences satisfy $v_a\ge u_a+h_{t,a}$.  Centering gives
\begin{equation}
u^\top v\ge Q_{t-1}+u^\top h_t=Q_{t-1}.
\label{eq:app-rm-angle}
\end{equation}
Thus $Q_t\ge Q_{t-1}$ by Cauchy--Schwarz.  The scalar inequality
$[(a+b)_+]^2\le[a_+]^2+2a_+b+b^2$ also gives
$Q_t\le Q_{t-1}+\norm{h_t}_2^2$.

Normalize $u$ and $v$ in $\ell_2$.  Equation~\ref{eq:app-rm-angle} and
$2(1-r^{-1/2})\le\log r$ for $r\ge1$ imply
\begin{equation}
\norm{\frac{u}{\norm{u}_2}-\frac{v}{\norm{v}_2}}_2^2
\le\log\frac{Q_t}{Q_{t-1}}.
\end{equation}
For nonnegative unit $\ell_2$ vectors, subsequent $\ell_1$ normalization is
$2\sqrt d$-Lipschitz.  The squared policy toll is therefore at most
$4d\log(Q_t/Q_{t-1})$ whenever $Q_{t-1}>0$.

Before $\tau$, every positive part has zero mass, so the fixed fallback makes
all those policies identical.  The move at $\tau$ has $\ell_1$ length at
most two, the diameter of the probability simplex.  Put $N=T-\tau$.  On the
$N$ later transitions, every $Q_t$ is positive and
\begin{align}
\sum_{t=\tau+1}^{T}\log\frac{Q_t}{Q_{t-1}}
&=\log\frac{Q_T}{q},\nonumber
\left(\sum_{t=\tau+1}^{T}\norm{p_{t+1}-p_t}_1\right)^2
&\le N\sum_{t=\tau+1}^{T}\norm{p_{t+1}-p_t}_1^2
\le4dN\log\frac{Q_T}{q}.
\label{eq:app-rm-temporal-telescope}
\end{align}
The first line is an exact adjacent-log telescope and the second is temporal
Cauchy--Schwarz followed by the one-step toll.  Iterating
$Q_t\le Q_{t-1}+\norm{h_t}_2^2$ from the activated state gives
$Q_T\le q+NG^2$.  Since $N\le T$ and the logarithm is increasing,
\[
2+2\sqrt{dN\log(Q_T/q)}
\le2+2\sqrt{dT\log(1+TG^2/q)},
\]
which proves Equation~\ref{eq:app-rm-path-bound}.  If a row never activates,
every policy is the same uniform fallback and its contribution is zero.  For
a fixed infinite execution, each activated row has one fixed $q>0$; summing
the displayed bound over the fixed finite row registry proves the global
$O_e(\sqrt{T\log(T+1)})$ rate. \hfill$\square$

\citet{burch2019revisiting} established the alternating correction below and
removed it under exact feedback through a monotonicity argument.  We restate
the correction in our notation and extend its use to \emph{partial} public
feedback, where realized policy motion supplies an appendix structural charge.
Their sign is recovered at
complete coverage through Lemma~\ref{lem:app-global-monotonicity}.

\begin{lemma}[Native alternating bridge; after \citealp{burch2019revisiting}]
\label{lem:app-alternating-bridge}
Let
\begin{equation}
L_i^T=\sum_{I\in\mathcal I_i}
\left[\max_a\sum_{t=1}^T g_{i,I,t}(a)\right]_+
\end{equation}
be exact target local regret, and let
$\bar x=T^{-1}\sum_t x_t$, $\bar y=T^{-1}\sum_t y_t$.  Define
\begin{equation}
C_T=\max_{y}\sum_{t=1}^T u(x_{t+1}-x_t,y)
-\sum_{t=1}^T u(x_{t+1}-x_t,y_t).
\label{eq:app-alternating-correction}
\end{equation}
Then
\begin{equation}
T\operatorname{Gap}(\bar x,\bar y)
\le L_0^T+L_1^T+C_T
\le L_0^T+L_1^T+[C_T]_+,
\label{eq:app-alternating-gap}
\end{equation}
and a finite game-dependent constant $L_{\rm alt}$ satisfies
$[C_T]_+\le L_{\rm alt}P_T$.
\end{lemma}

\paragraph{Proof.}
The perfect-recall local-to-external reduction, applied at the two actual
event origins, gives
\begin{align}
\max_x\sum_t\{u(x,y_t)-u(x_t,y_t)\}&\le L_0^T,\\
\max_y\sum_t\{u(x_{t+1},y_t)-u(x_{t+1},y)\}&\le L_1^T.
\end{align}
Adding and subtracting $x_{t+1}-x_t$ in the second inequality gives
Equation~\ref{eq:app-alternating-gap}.  This supplies the alternating-origin
correction absent from the simultaneous-update argument.  Finiteness gives a constant $L_u^0$
such that $|u(x'-x,y)|$ is at most $L_u^0$ times player zero's intervening
behavior path.  Applying it to both terms in
Equation~\ref{eq:app-alternating-correction} gives
$[C_T]_+\le2L_u^0P_{0,T}\le2L_u^0P_T$; take
$L_{\rm alt}=2L_u^0$. \hfill$\square$

If all terminal payoffs lie in $[u_{\min},u_{\max}]$, put
$\Delta_u=u_{\max}-u_{\min}$.  Total variation at one changed behavior row
and root-to-leaf telescoping permit $L_u^0=\Delta_u/2$.  The same bridge
identity also gives the fully calibrated trace bound
\begin{equation}
[C_T]_+\le
\min\left\{
\Delta_uP_{0,T},\;
\Delta_u\left(1+\frac{P_{0,T}}2\right),\;
\Delta_u+\sum_{t=1}^T
 [u(x_t,y_t)-u(x_{t+1},y_t)]_+
\right\}.
\label{eq:app-calibrated-alternating-bridge}
\end{equation}
Indeed the first term bounds both summands in
Equation~\ref{eq:app-alternating-correction} by rowwise total variation; the
second bounds its fixed-$y$ telescoping endpoint by $\Delta_u$.  For the
third, that endpoint has the same span bound and the remaining negative sum
is at most the sum of its positive realized target-payoff decreases.  This is
an a-posteriori target-game bound rather than a learner monotonicity claim.

\begin{lemma}[Weighted epoch-start output is a path coreset]
\label{lem:app-coverage-coreset}
Additionally assume that every relevant lane has one common epoch length $m$
and both players' epoch starts align.  Let $z_{i,t}$ be player $i$'s pre-event
realization plan, fix nonnegative nondecreasing weights $w_t$ with
$W_T=\sum_{t\le T}w_t>0$, and partition the stopped horizon into epochs
$E_k=\{km+1,\ldots,\min((k+1)m,T)\}$.  With $s_k=km+1$ and
$\Omega_{k,T}=\sum_{t\in E_k}w_t$, define
\begin{equation}
\bar z_i^w=\frac1{W_T}\sum_{t=1}^Tw_tz_{i,t},\qquad
\widetilde z_i^w=\frac1{W_T}\sum_k\Omega_{k,T}z_{i,s_k}.
\end{equation}
There is a finite realization-map constant $L_Q^i$ such that
\begin{equation}
\norm{\widetilde z_i^w-\bar z_i^w}_1
\le\frac{L_Q^i\kappa_T(m)P_{i,T}}{W_T}
\le\frac{L_Q^iw_T(m-1)P_{i,T}}{W_T},
\label{eq:app-coverage-coreset}
\end{equation}
where
$\kappa_T(m)=\max(\{0\}\cup\{\sum_{t=q+1}^{\max E_k}w_t:
s_k\le q<\max E_k\})$.  At $m=1$ the two outputs are identical at every
horizon.  A stopped final epoch needs no diameter penalty because its retained
start receives its actual accumulated mass $\Omega_{k,T}$.
\end{lemma}

\paragraph{Proof.}
Within epoch $E_k$, expand
$\sum_{t\in E_k}w_t(z_{i,s_k}-z_{i,t})$ by telescoping along its intervening
own-player transitions.  The transition after $q$ receives exactly the tail
coefficient $\sum_{t=q+1}^{\max E_k}w_t$, at most $\kappa_T(m)$ and then at
most $(m-1)w_T$.  Sum the disjoint within-epoch transitions, apply
realization-map Lipschitzness, and divide by $W_T$.  Actual epoch masses sum
to $W_T$, proving feasibility and the stopped-horizon statement.  Player
one's observer call follows player zero's update, but its own realization plan
remains $y_t$; thus both players use the aligned logical starts. \hfill$\square$

\subsection{Weighted target transfer and additive RM+}
\label{app:weighted-rmplus}

Fix one predetermined deterministic clock
$0\le w_1\le\cdots$, shared by the two player events in logical round $t$ and
independent of public order and learner history, with
$W_T=\sum_{t\le T}w_t>0$.  Each stream has a strictly increasing pullback to
the global logical clock and obeys Equation~\ref{eq:app-stream-incidence}; a
physically missing event contributes zero target/delivered error.  For the native pre/pre observer, define
$(\bar x_T^w,\bar y_T^w)=W_T^{-1}\sum_t w_t(x_t,y_t)$ in realization-plan
space; behaviorally this is the standard own-reach weighted average.

\begin{lemma}[Weighted RM+ envelope]
\label{lem:app-weighted-rmplus-envelope}
Let $q_0=0$, $q_t=[q_{t-1}+h_t]_+$,
$p_t=\operatorname{RM}(q_{t-1})$, $\langle p_t,h_t\rangle=0$, and
$\norm{h_t}_2\le G$.  For nondecreasing nonnegative $w_t$,
\begin{equation}
 \left[\max_a\sum_{t=1}^T w_th_t(a)\right]_+
 \le w_T\norm{q_T}_\infty\le w_TG\sqrt T .
\label{eq:app-weighted-rmplus-envelope}
\end{equation}
\end{lemma}

\paragraph{Proof.}
The coordinatewise clipping correction
$c_t=q_t-q_{t-1}-h_t$ is nonnegative.  Summation by parts gives
\begin{equation}
 \sum_{t=1}^T w_th_t
 =w_Tq_T-\sum_{t=1}^{T-1}(w_{t+1}-w_t)q_t
       -\sum_{t=1}^T w_tc_t
 \le w_Tq_T .
\label{eq:app-weighted-rmplus-telescope}
\end{equation}
The centered-potential calculation gives
$\norm{q_T}_2^2\le\sum_t\norm{h_t}_2^2\le TG^2$. \hfill$\square$

The RM+ lemma will later bound delivered weighted regret.  First transfer an
arbitrary discrepancy-controlled stream.  Pull the common clock back through
the strictly increasing map $\tau_s$ and write $u_n=w_{\tau_s(n)}$.  For one
label, let
\begin{equation}
B_{q,j}=\sum_{n=1}^q u_na_{n,j}
=u_qA_{q,j}-\sum_{n=1}^{q-1}(u_{n+1}-u_n)A_{n,j}.
\label{eq:app-weighted-prefix-discrepancy}
\end{equation}
Because $u_n$ is nondecreasing,
$|B_{q,j}|\le(2u_q-u_1)D_T\le2w_TD_T$, including every stopped horizon.

\begin{lemma}[Global weighted discrepancy--path transfer]
\label{lem:app-global-weighted-transfer}
Pull the common clock back to every finite asynchronous player/lane stream
satisfying Equation~\ref{eq:app-stream-incidence}.
For each row $\rho$ and stream $s$, choose before execution structural bounds
\[
\sup_\sigma\sum_j\norm{X_{\rho,s,j}(\sigma)}_\infty\le M_{\rho,s},\qquad
\sum_j\norm{X_{\rho,s,j}(\sigma')-X_{\rho,s,j}(\sigma)}_\infty
\le L_{\rho,s}d_1(\sigma',\sigma).
\]
With $C_0=\sum_{\rho,s}M_{\rho,s}$ and
$C_X=\sum_{\rho,s}L_{\rho,s}$,
\begin{equation}
\sum_\rho\norm{\sum_{t=1}^T w_t
 (h_{\rho,t}-g_{\rho,t})}_\infty
\le2C_0w_TD_T+2C_Xw_TD_TP_T.
\label{eq:app-global-weighted-transfer}
\end{equation}
\end{lemma}

\paragraph{Proof.}
By the incidence identity, the aggregate sum on the left of
Equation~\ref{eq:app-global-weighted-transfer} equals the sum of the weighted
component streams incident to row $\rho$.  Apply component summation by parts
using $B_{q,j}$.  Its endpoint is at most
$2w_TD_TM_{\rho,s}$ and its variation at most
$2w_TD_TL_{\rho,s}$ times consecutive-origin motion.  Gaps in $\tau_s$ are
harmless: these origins delimit disjoint global path segments within a stream;
finite cross-stream reuse is absorbed by $C_X$.  A stream with no event
contributes zero.  Summing proves
Equation~\ref{eq:app-global-weighted-transfer}. \hfill$\square$

Exact balance admits sharper constants.  In one complete $m$-batch epoch
beginning after event $t_0$, abbreviate $w_r=w_{t_0+r}$.  If batch $j$ appears
at position $s=p(j)$, put
\begin{equation}
 b_{r,j}=w_r(m\ind{s=r}-1),\qquad
 B_{r,j}=\sum_{q=1}^r b_{q,j}.
\end{equation}
Then
\begin{equation}
 \sum_{r=1}^m b_{r,j}X_{r,j}
 =D_jX_{m,j}-\sum_{r=1}^{m-1}B_{r,j}(X_{r+1,j}-X_{r,j}),
 \quad D_j=mw_s-\sum_{r=1}^m w_r,
\label{eq:app-weighted-balance-identity}
\end{equation}
with sharp endpoint radius
\begin{equation}
 \beta_k=\max\left\{
 \sum_{r=1}^m(w_r-w_1),
 \sum_{r=1}^m(w_m-w_r)\right\},
 \qquad |B_{r,j}|\le mw_m .
\label{eq:app-weighted-backlog-radius}
\end{equation}
Across contiguous complete epochs,
$\sum_k\beta_k\le(m-1)(w_T-w_1)$; a stopped final epoch contributes at most
$mw_TX^{\max}$ per fixed row/lane.  These yield the exact linear, quadratic,
and delayed-linear backlog formulas below and sharpen the corresponding
specialization of the weighted master theorem.

\begin{lemma}[Weighted perfect-recall reduction]
\label{lem:app-weighted-perfect-recall}
For common nonnegative logical-round weights, define
\begin{equation}
L_i^{w,T}=\sum_{I\in\mathcal I_i}
 \left[\max_a\sum_{t=1}^T w_t g_{i,I,t}(a)\right]_+ .
\label{eq:app-weighted-target-local-regret}
\end{equation}
At the native event origins,
\begin{align}
\max_x\sum_t w_t\{u(x,y_t)-u(x_t,y_t)\}&\le L_0^{w,T},\nonumber\\
\max_y\sum_t w_t\{u(x_{t+1},y_t)-u(x_{t+1},y)\}&\le L_1^{w,T}.
\label{eq:app-weighted-local-to-external}
\end{align}
\end{lemma}

\paragraph{Proof.}
We give the player-zero argument; player one is identical at its post/pre
origins with payoff $-u$.  Fix any valid behavioral comparator $x'$.  The
one-round perfect-recall policy-difference identity is
\begin{align}
u(x',y_t)-u(x_t,y_t)
=\sum_{I\in\mathcal I_0}\pi_0^{x'}(I)
 \sum_{a\in A(I)}x'(I,a)g_{0,I,t}(a).
\label{eq:app-one-round-comparator-decomposition}
\end{align}
To see this, telescope the terminal realization probability from $x_t$ to
the profile that uses $x'$ at every player-zero row, assign each changed
decision occurrence to its information set, and factor the comparator's own
prefix reach.  Perfect recall makes that prefix identical at all histories in
the information set; opponent and chance reach remain in the unnormalized
counterfactual row.  Centering removes the old-policy baseline because both
$x_t(I)$ and $x'(I)$ are distributions.

Multiply Equation~\ref{eq:app-one-round-comparator-decomposition} by $w_t$
and interchange the two finite sums.  For the cumulative row
$G_I=\sum_tw_tg_{0,I,t}$,
\[
\sum_a x'(I,a)G_I(a)\le\max_aG_I(a)
\le[\max_aG_I(a)]_+.
\]
Moreover $0\le\pi_0^{x'}(I)\le1$, so multiplying the nonnegative final
envelope by this reach cannot increase it.  Summing $I$ proves the first
inequality of Equation~\ref{eq:app-weighted-local-to-external}; taking the
supremum over valid $x'$ is legitimate because the uniform behavioral
comparator makes the comparator domain nonempty and the displayed local sum
is a common upper bound.  Apply the same argument
to player one at $(x_{t+1},y_t)$.  No stochastic assumption is used.
\hfill$\square$

For the pre/pre average, the native alternating bridge is
\begin{align}
 W_T\operatorname{Gap}(\bar x_T^w,\bar y_T^w)
 &\le L_0^{w,T}+L_1^{w,T}+C_T^w,\nonumber\\
 C_T^w
 &=\max_y\sum_t w_tu(x_{t+1}-x_t,y)
   -\sum_t w_tu(x_{t+1}-x_t,y_t),
\label{eq:app-weighted-alternating-bridge}
\end{align}
and finiteness gives $[C_T^w]_+\le L_{\rm alt}w_TP_T$.  Combining these
facts gives the recurrence-agnostic bound.

\begin{theorem}[Weighted master target-transfer bound]
\label{thm:app-weighted-master}
Define realized delivered weighted local regret
\[
\widehat{\mathcal R}_T^w
=\sum_{i,I}\left[\max_a\sum_{t=1}^Tw_th_{i,I,t}(a)\right]_+ .
\]
Then structural $C_0,C_X,L_{\rm alt}$, independent of realized public order
and behavior path, satisfy
\begin{equation}
\operatorname{expl}(\bar x_T^w,\bar y_T^w)
\le
\frac{\widehat{\mathcal R}_T^w+2C_0w_TD_T
+(2C_XD_T+L_{\rm alt})w_TP_T}{2W_T}.
\label{eq:app-weighted-master}
\end{equation}
\end{theorem}

\paragraph{Proof.}
Weighted exact target local regret is at most delivered weighted local regret
plus the rowwise norm in
Equation~\ref{eq:app-global-weighted-transfer}.  Apply
Lemma~\ref{lem:app-weighted-perfect-recall} and
Equation~\ref{eq:app-weighted-alternating-bridge}, then divide by $2W_T$.
\hfill$\square$

For additive RM+,
Lemma~\ref{lem:app-weighted-rmplus-envelope} gives
$\widehat{\mathcal R}_T^w\le C_Rw_T\sqrt T$.  Substitution into
Theorem~\ref{thm:app-weighted-master}, together with
Lemma~\ref{lem:app-rm-log-path}, gives
$O_e(\sqrt{\log(T+1)/T})$ exploitability whenever $D_T$ is bounded and
$w_T/W_T=O(1/T)$.

The observer-drift terms are explicit.  At a complete horizon $T=mK$,
\begin{align}
 w_t=t:\qquad &W_T=\frac{T(T+1)}2,&
 \sum_k\beta_k&=\frac{(m-1)T}2,\label{eq:app-linear-backlog}\\
 w_t=t^2:\qquad &W_T=\frac{T(T+1)(2T+1)}6,&
 \sum_k\beta_k&=\frac{(m-1)T(3T+m+1)}6.
\label{eq:app-quadratic-backlog}
\end{align}
For delayed-linear $w_t=(t-d)_+$, let integer $0\le d<T$,
$n=T-d$, and $c=d\bmod m$; then
\begin{equation}
 W_T=\frac{n(n+1)}2,\qquad
 \sum_k\beta_k=\frac{n(m-1)+c(m-c)}2.
\label{eq:app-delayed-linear-backlog}
\end{equation}
Thus delayed linear has the pointwise
$O_e(\sqrt{\log(T+1)/T})$ structural rate when $T-d=\Omega(T)$; without that
condition the bound is
$O_e(\sqrt{T\log(T+1)}/(T-d)+1/(T-d))$.

The original CFR+ delayed observer averages the between/pre pair
$(x_{t+1},y_t)$, rather than the native solver's pre/pre pair.  Define
$\bar x_{T,\mathrm{post}}^w=W_T^{-1}\sum_t w_tx_{t+1}$ in realization-plan
space.  Repeating the weighted local-to-global reduction gives the sharper bridge
\begin{equation}
 W_T\operatorname{Gap}(\bar x_{T,\mathrm{post}}^w,\bar y_T^w)
 \le L_0^{w,T}+L_1^{w,T}
 -\sum_t w_tu(x_{t+1}-x_t,y_t).
\label{eq:app-weighted-between-pre-bridge}
\end{equation}
For partial balanced feedback, finiteness and \(w_t\le w_T\) give
\begin{equation}
\left|\sum_t w_tu(x_{t+1}-x_t,y_t)\right|
\le L_{\rm alt}w_TP_T ,
\label{eq:app-weighted-between-pre-partial}
\end{equation}
so this observer has the same partial-feedback rate.  At exact target
feedback, Lemma~\ref{lem:app-global-monotonicity} gives the sharper sign
$u(x_{t+1}-x_t,y_t)\ge0$, hence
$-\sum_t w_tu(x_{t+1}-x_t,y_t)\le0$.  For pre/pre output, define
$\Omega_0=\max_{x,x',y}\{u(x,y)-u(x',y)\}$.  Weighted Abel summation bounds
the exact alternating endpoint by $\Omega_0w_T$.  Hence at complete coverage
\begin{equation}
 \operatorname{expl}(\bar x_T^w,\bar y_T^w)
 \le \frac{C_Rw_T\sqrt T+\Omega_0w_T}{2W_T},
\label{eq:app-weighted-exact-boundary}
\end{equation}
while the between/pre bridge drops the $\Omega_0$ term.  These statements
cover the original delayed-linear and native linear/quadratic additive CFR+
observer semantics.  The native partial-public CFR+ experiment instantiates
the same recurrence and uses Lemma~\ref{lem:app-coverage-coreset} to compress
its output events.  Discounted and predictive recurrences require their own
recurrence-specific transfer.

\subsection{Component-resolved trace certificate}
\label{app:deadline-variation}

Fix an actual coefficient stream $s=(i,\ell)$ and let
$\mathcal R_s$ be the rows governed by its one physical order and let $J_s$
be its semantic public registry.  For every local epoch $k\ge0$, the realized
ordered batches
$(\mathcal B_{s,k,1},\ldots,\mathcal B_{s,k,m_s})$ partition $J_s$; the
partition may depend on $k$.  Let $p_{s,k}(\lambda)$ be the unique position
whose batch contains semantic label $\lambda$.  For position
$1\le r\le m_s$, put
$n_s(k,r)=km_s+r$ and $t_s(k,r)=\tau_s(n_s(k,r))$.  The first is a local
stream index; the second is its actual global owner round.  Define
\begin{equation}
X_{\rho,s,k,r,\lambda}
=X_{\rho,\ell,n_s(k,r),\lambda},
\label{eq:app-batch-component}
\end{equation}
with every component evaluated at the immutable consuming profile in global
round $t_s(k,r)$.
Exact additive ownership and same-origin centering give
\begin{equation}
h_{\rho,s,k,r}-g_{\rho,s,k,r}
=\sum_{\lambda\in J_s}
\left(m_s\ind{p_{s,k}(\lambda)=r}-1\right)
X_{\rho,s,k,r,\lambda}.
\label{eq:app-batch-semantic-interface}
\end{equation}
Write this stream contribution as $e_{\rho,s,k,r}$ and put
$Z_{\rho,s,k}=\sum_{r=1}^{m_s}e_{\rho,s,k,r}$.

\begin{lemma}[Exact local deadline--variation identity]
\label{lem:app-deadline-variation}
For every realized complete epoch, define
\begin{equation}
a_{r,\lambda}=m_s\ind{p_{s,k}(\lambda)=r}-1,\qquad
A_{r,\lambda}=\sum_{q=1}^r a_{q,\lambda}
=m_s\ind{p_{s,k}(\lambda)\le r}-r .
\end{equation}
Then
\begin{equation}
\sum_{r=1}^{m_s}\sum_{\lambda\in J_s}
a_{r,\lambda}X_{\rho,s,k,r,\lambda}
=-\sum_{\lambda\in J_s}\sum_{r=1}^{m_s-1}
A_{r,\lambda}(X_{\rho,s,k,r+1,\lambda}
-X_{\rho,s,k,r,\lambda}).
\label{eq:app-deadline-variation-identity}
\end{equation}
Consequently the infinity norm of the signed left side is at most
\begin{equation}
\mathcal E_{\rho,s,k}^\infty
=\sum_{\lambda\in J_s}\sum_{r=1}^{m_s-1}|A_{r,\lambda}|
\norm{X_{\rho,s,k,r+1,\lambda}
-X_{\rho,s,k,r,\lambda}}_\infty .
\label{eq:app-deadline-variation-debit}
\end{equation}
\end{lemma}

\paragraph{Proof.}
For fixed $\lambda$, $a_{r,\lambda}=A_{r,\lambda}-A_{r-1,\lambda}$
with $A_{0,\lambda}=0$.
Discrete summation by parts gives
\begin{equation}
\sum_{r=1}^{m_s}a_{r,\lambda}X_{r,\lambda}
=A_{m_s,\lambda}X_{m_s,\lambda}
-\sum_{r=1}^{m_s-1}A_{r,\lambda}
(X_{r+1,\lambda}-X_{r,\lambda}).
\end{equation}
Exact coverage makes $A_{m_s,\lambda}=0$.  Sum over semantic labels and apply the
infinity-norm triangle inequality. \hfill$\square$

Call $T$ locally epoch-closed when $m_s\mid n_s(T)$ for every stream.
Partition the streams incident to each row into declared
\emph{shared-coefficient groups}: streams in one group have a common epoch
length, owner-round incidence, and epoch-local realized position map under a
declared fixed identification of their semantic registries.  Sum their
component arrays under that identification before applying
Equation~\ref{eq:app-deadline-variation-debit}, and let
$\mathcal G_\rho$ denote the partition for row $\rho$.  Let
$\mathcal E_{Q,T}^\infty$ denote the resulting sum over rows, groups, and
complete local epochs; $\mathcal E_T^\infty$ denotes its singleton-group
counterpart.  As in the structural transfer argument above,
$\widehat{\mathcal R}_T$ is delivered positive local regret summed over rows.

\begin{theorem}[Asynchronous component-resolved finite bound]
\label{thm:deadline-weighted-refinement}
At every locally epoch-closed horizon $T\ge1$ and for every declared
shared-coefficient partition,
\begin{equation}
\operatorname{expl}(\bar x_T,\bar y_T)
\le\frac{\widehat{\mathcal R}_T+\mathcal E_{Q,T}^\infty
+L_{\rm alt}P_T}{2T}.
\label{eq:deadline-weighted-bound}
\end{equation}
The debit and path are realized trace quantities; the denominator counts
native alternating rounds while closure is tested on each local event clock.
\end{theorem}

\paragraph{Proof of Theorem~\ref{thm:deadline-weighted-refinement}.}
Let $\mathcal S_\rho$ be the streams contributing to row $\rho$, equivalently
$s\in\mathcal S_\rho$ if and only if $\rho\in\mathcal R_s$.  Assume
$m_s\mid n_s(T)$ for every declared stream and put
$K_s(T)=n_s(T)/m_s$.  Summing the incidence identity over global rounds and
grouping consecutive \emph{local} events into complete epochs gives
\begin{align}
\sum_{t=1}^T(h_{\rho,t}-g_{\rho,t})
&=\sum_{s\in\mathcal S_\rho}\sum_{n=1}^{n_s(T)}
  \sum_{\lambda\in J_s}a_{s,n,\lambda}
  X_{\rho,s,n,\lambda}\nonumber\\
&=\sum_{s\in\mathcal S_\rho}\sum_{k=0}^{K_s(T)-1}Z_{\rho,s,k}.
\label{eq:app-deadline-incidence-pullback}
\end{align}
Streams inside one declared shared-coefficient group have a common epoch
length, epoch-local position map under their declared label identification,
and owner-round incidence.  Write $K_G(T)$ for their common number of complete
epochs.  Their epoch sums share one coefficient array $a_{r,\lambda}$; because
Equation~\ref{eq:app-deadline-variation-identity} is linear in the component
array, it applies verbatim to the group-summed components, and bounding each
group after summation is tighter than bounding its members separately.  Thus
the local identity and triangle inequality imply
\begin{equation}
\sum_\rho\norm{\sum_{t=1}^T(h_{\rho,t}-g_{\rho,t})}_\infty
\le
\sum_\rho\sum_{G\in\mathcal G_\rho}
\sum_{k=0}^{K_G(T)-1}\mathcal E_{\rho,G,k}^\infty
=\mathcal E_{Q,T}^\infty
\le\mathcal E_T^\infty .
\label{eq:app-deadline-row-transfer}
\end{equation}
Target positive local regret is at most delivered positive local regret plus
the corresponding row infinity norm.  Lemma~\ref{lem:app-alternating-bridge}
adds $L_{\rm alt}P_T$.  Divide by $T$ and by two under the paper's
exploitability convention.  This proves Equation~\ref{eq:deadline-weighted-bound};
additive signed RM/RM+ further gives
$\widehat{\mathcal R}_T\le C_R\sqrt T$.  Gaps in $\tau_s$ affect which
profiles enter the debit but not the global alternating bridge; they mean zero
error from stream $s$, not that other streams or exact remainder terms vanish.

For a residual local prefix $1\le q<m_s$, the exact identity is
\begin{equation}
\sum_{r=1}^{q}\sum_{\lambda\in J_s}a_{r,\lambda}X_{r,\lambda}
=\sum_{\lambda\in J_s}A_{q,\lambda}X_{q,\lambda}
-\sum_{\lambda\in J_s}\sum_{r=1}^{q-1}A_{r,\lambda}
(X_{r+1,\lambda}-X_{r,\lambda}).
\label{eq:app-deadline-residual-prefix}
\end{equation}
Its debit includes both the endpoint
$\sum_{\lambda\in J_s}|A_{q,\lambda}|
\norm{X_{q,\lambda}}_\infty$ and the displayed truncated variation;
at $q=0$ both are zero.  Theorem~\ref{thm:deadline-weighted-refinement} is
unweighted.  The following weighted identity retains the generally nonzero
epoch endpoint instead of treating exact unweighted balance as weighted
cancellation.

For one complete epoch, reinstate
$a_{s,k,r,\lambda}=m_s\ind{p_{s,k}(\lambda)=r}-1$ and define
\begin{align}
u_{s,k,r}&=w_{t_s(k,r)},&
b_{s,k,r,\lambda}&=u_{s,k,r}a_{s,k,r,\lambda},&
B_{s,k,r,\lambda}&=\sum_{q=1}^r b_{s,k,q,\lambda}.
\label{eq:app-weighted-component-prefix}
\end{align}

\begin{lemma}[Exact weighted local deadline--variation identity]
\label{lem:app-weighted-deadline-variation}
For every realized complete epoch,
\begin{align}
Z_{\rho,s,k}^w
&=\sum_{r=1}^{m_s}\sum_{\lambda\in J_s}
b_{s,k,r,\lambda}X_{\rho,s,k,r,\lambda}\nonumber\\
&=\sum_{\lambda\in J_s}B_{s,k,m_s,\lambda}
X_{\rho,s,k,m_s,\lambda}
-\sum_{\lambda\in J_s}\sum_{r=1}^{m_s-1}
B_{s,k,r,\lambda}
\bigl(X_{\rho,s,k,r+1,\lambda}
-X_{\rho,s,k,r,\lambda}\bigr).
\label{eq:app-weighted-deadline-identity}
\end{align}
Consequently, $\norm{Z_{\rho,s,k}^w}_\infty$ is at most
\begin{align}
\mathcal E_{\rho,s,k}^{w,\infty}
={}&\sum_{\lambda\in J_s}|B_{s,k,m_s,\lambda}|
\norm{X_{\rho,s,k,m_s,\lambda}}_\infty\nonumber\\
&+\sum_{\lambda\in J_s}\sum_{r=1}^{m_s-1}
|B_{s,k,r,\lambda}|
\norm{X_{\rho,s,k,r+1,\lambda}
-X_{\rho,s,k,r,\lambda}}_\infty.
\label{eq:app-weighted-deadline-debit}
\end{align}
Under a unit clock, the endpoint vanishes and this reduces to
Lemma~\ref{lem:app-deadline-variation}.
\end{lemma}

\paragraph{Proof.}
For each semantic label, apply discrete summation by parts to
$b_{r,\lambda}=B_{r,\lambda}-B_{r-1,\lambda}$, retain the endpoint
$B_{m_s,\lambda}X_{m_s,\lambda}$, and then sum over labels.  The norm bound
is the triangle inequality.  If $u_{s,k,r}=1$, then
$B_{s,k,m_s,\lambda}=A_{s,k,m_s,\lambda}=0$. \hfill$\square$

At a locally epoch-closed horizon, sum
Equation~\ref{eq:app-weighted-deadline-debit} after the same declared
shared-coefficient grouping as in the unweighted theorem; denote the result
by $\mathcal E_{Q,T}^{w,\infty}$.  Let $\mathcal A_T^w$ be any nonnegative
certified upper bound on the positive part of the weighted alternating term
$C_T^w$ in Equation~\ref{eq:app-weighted-alternating-bridge}.  The native
certificate uses the minimum of its path, endpoint--path, and realized
payoff-decrease upper bounds.

\begin{theorem}[Weighted component-resolved finite certificate]
\label{thm:app-weighted-component-certificate}
At every locally epoch-closed horizon and for every declared
shared-coefficient partition,
\begin{equation}
\operatorname{expl}(\bar x_T^w,\bar y_T^w)
\le
\frac{\widehat{\mathcal R}_T^w
+\mathcal E_{Q,T}^{w,\infty}+\mathcal A_T^w}{2W_T}.
\label{eq:app-weighted-component-certificate}
\end{equation}
For simultaneous updates the alternating term is absent, so one may take
$\mathcal A_T^w=0$.
\end{theorem}

\paragraph{Proof.}
Pull $w_t(h_{\rho,t}-g_{\rho,t})$ through every stream incidence map.  Group
the resulting complete local epochs and apply
Lemma~\ref{lem:app-weighted-deadline-variation} before taking row norms.
Target weighted positive local regret is at most delivered weighted positive
local regret plus the resulting rowwise debit.  Apply
Lemma~\ref{lem:app-weighted-perfect-recall} and the weighted alternating
bridge, then divide the Nash gap by $2W_T$. \hfill$\square$

\paragraph{Weighted epoch-start output.}
Let $\widetilde\sigma_T^w=(\widetilde x_T^w,\widetilde y_T^w)$ be
realization-equivalent to the pair of aligned weighted epoch-start coresets
when one common epoch length $m$ exists.
Because the finite realization polytopes are compact and the payoff is
bilinear, fix finite coordinatewise target-gap Lipschitz constants
$\Lambda_0,\Lambda_1$ such that, for all feasible realization plans,
\begin{equation}
\left|\operatorname{Gap}(x',y')-\operatorname{Gap}(x,y)\right|
\le \Lambda_0\norm{x'-x}_1+\Lambda_1\norm{y'-y}_1.
\label{eq:app-target-gap-lipschitz}
\end{equation}
The maximum and minimum over the finite pure-strategy sets preserve the
Lipschitz envelopes of their linear payoff functions.  Exploitability is
one half of this gap.

\begin{corollary}[Weighted epoch-start path coreset]
\label{cor:app-epoch-start}
At every stopped horizon,
\begin{equation}
\operatorname{expl}(\widetilde\sigma_T^w)
\le\operatorname{expl}(\bar x_T^w,\bar y_T^w)
+\frac{\kappa_T(m)}{2W_T}
(\Lambda_0L_Q^0P_{0,T}+\Lambda_1L_Q^1P_{1,T}).
\label{eq:app-epoch-start-bound}
\end{equation}
For quadratic $w_t=t^2$, $w_T/W_T\le3/T$; hence fixed $m$ preserves the
$O_e(\sqrt{\log(T+1)/T})$ additive-RM+ structural rate in
Corollary~\ref{cor:weighted-balanced-rmplus}.
\end{corollary}

This follows directly from Lemma~\ref{lem:app-coverage-coreset} and
Equation~\ref{eq:app-target-gap-lipschitz}.

The structural constants in Corollary~\ref{cor:app-epoch-start} are useful
for asymptotic rates.  A finite replay can instead certify the realized
coreset correction without estimating those constants.  Let
$\mathcal X,\mathcal Y$ denote the two feasible realization-plan polytopes,
extend the bilinear payoff $u$ linearly to signed realization plans, and define
\begin{align}
S_{0,T}^w
&=\left[\max_{x\in\mathcal X}
u(x,\widetilde y_T^w-\bar y_T^w)\right]_+,
&
S_{1,T}^w
&=\left[\max_{y\in\mathcal Y}
u(\bar x_T^w-\widetilde x_T^w,y)\right]_+ .
\label{eq:app-signed-coreset-supports}
\end{align}

\begin{lemma}[Exact signed-support coreset transfer]
\label{lem:app-signed-coreset-transfer}
For every pair of dense and coreset realization plans,
\begin{equation}
\operatorname{expl}(\widetilde x_T^w,\widetilde y_T^w)
\le
\operatorname{expl}(\bar x_T^w,\bar y_T^w)
+\frac{S_{0,T}^w+S_{1,T}^w}{2}.
\label{eq:app-signed-coreset-transfer}
\end{equation}
Both support values are finite ordinary best-response optimizations with a
signed opponent realization vector.
\end{lemma}

\paragraph{Proof.}
For any functions $f,g$ on a common set,
$\max f-\max g\le\max(f-g)$.  Apply this once to the row best-response term:
\[
\max_xu(x,\widetilde y_T^w)-\max_xu(x,\bar y_T^w)
\le\max_xu(x,\widetilde y_T^w-\bar y_T^w),
\]
and once to the column term after writing its contribution as
$\max_y[-u(\widetilde x_T^w,y)]-\max_y[-u(\bar x_T^w,y)]$.
The latter is at most
$\max_yu(\bar x_T^w-\widetilde x_T^w,y)$.  Replacing either support by its
positive part preserves the upper bound.  Add the two inequalities and
divide the Nash gap by two. \hfill$\square$

\begin{corollary}[Complete weighted RM+ output certificate]
\label{cor:app-complete-weighted-output-certificate}
Under the premises of
Theorem~\ref{thm:app-weighted-component-certificate} and an aligned
epoch-start output,
\begin{equation}
\operatorname{expl}(\widetilde x_T^w,\widetilde y_T^w)
\le
\frac{\widehat{\mathcal R}_T^w
+\mathcal E_{Q,T}^{w,\infty}+\mathcal A_T^w}{2W_T}
+\frac{S_{0,T}^w+S_{1,T}^w}{2}.
\label{eq:app-complete-weighted-output-certificate}
\end{equation}
For additive RM+, Lemma~\ref{lem:app-weighted-rmplus-envelope} supplies the
delivered-regret envelope.  At $m=1$, both the quotient endpoint and the
signed-support correction vanish.
\end{corollary}

\begin{corollary}[Weighted transfer and epoch coreset]
\label{cor:coverage-clock-output}
\phantomsection\label{cor:weighted-balanced-rmplus}%
For predetermined $0\le w_1\le\cdots\le w_T$, let
$W_T=\sum_{t\le T}w_t>0$ and
$\widehat{\mathcal R}_T^w=\sum_\rho[\max_a\sum_{t\le T}
w_th_{\rho,t}(a)]_+$.  If all lanes share an aligned epoch length $m$, there
exists a finite game-dependent constant $C_{\rm cov}$ such that
\begin{align}
\operatorname{expl}(\widetilde x_T^w,\widetilde y_T^w)
&\le \operatorname{expl}(\bar x_T^w,\bar y_T^w)
+\frac{C_{\rm cov}w_T(m-1)P_T}{W_T},
\label{eq:main-coreset-output}\\
\operatorname{expl}(\bar x_T^w,\bar y_T^w)
&\le \frac{\widehat{\mathcal R}_T^w+2C_0w_TD_T
+(2C_XD_T+L_{\rm alt})w_TP_T}{2W_T}.
\label{eq:weighted-balanced-rmplus}
\end{align}
For additive RM+, $\widehat{\mathcal R}_T^w\le C_Rw_T\sqrt T$; exact balance,
$w_T/W_T=O(1/T)$, and Lemma~\ref{lem:app-rm-log-path} give
$O_e(\sqrt{\log(T+1)/T})$ for both outputs at fixed $m$.  At $m=1$, the two
outputs coincide.
\end{corollary}

\begin{lemma}[Player-zero delivered-payoff improvement]
\label{lem:app-global-monotonicity}
At one player-zero event, suppose all player-zero rows read one immutable
policy $\sigma$, all action values come from one common delivered payoff
functional $\widehat u$, and all additive RM or RM+ rows update
simultaneously, using the uniform distribution whenever a row has zero
positive regret mass.  Then
$\widehat u(\sigma^+,y)\ge\widehat u(\sigma,y)$.
\end{lemma}

\paragraph{Proof.}
For one row, set $z=R$, $p=\operatorname{RM}(z)$,
$h=v-\langle p,v\rangle\one$, $x=[z]_+$, and $y=[z+h]_+$.  Coordinatewise
monotonicity of positive part gives $(y_a-x_a)h_a\ge0$.  Since $x^Th=0$,
$y^Th\ge0$.  If $y$ has positive mass, normalization yields
$\langle\operatorname{RM}(z+h),h\rangle\ge0$.  If it has zero mass, then an
old positive-mass state is impossible: $x^Th=0$ and the projection
inequality force positive mass to persist.  Thus both old and new policies
use the same uniform fallback and the inner product is zero.  Since
$\langle p,h\rangle=0$, in every branch
$\langle\operatorname{RM}(z+h)-p,v\rangle\ge0$.  Applying positive part to
the stored RM+ state does not alter its regret-matching policy, so the same
argument covers RM+.

For the full perfect-recall policy, the exact performance-difference identity
is
\begin{equation}
\widehat u(\sigma^+,y)-\widehat u(\sigma,y)
=\sum_{I\in\mathcal I_i}\pi_i^{\sigma^+}(I)
\langle\sigma^+(I)-\sigma(I),v^\sigma(I,\cdot)\rangle.
\label{eq:app-performance-difference}
\end{equation}
To prove it, expand the
old/new terminal-probability difference as a telescoping sum over changed
player-$i$ decision occurrences, group occurrences by information set, and
factor their new-policy own prefixes.  Perfect recall makes that prefix
representative-independent and excludes a repeated copy of the same
information set below itself; opponent and chance reach remain in
$v^\sigma$.  This gives Equation~\ref{eq:app-performance-difference} without
normalizing unreachable rows.  Every reach is nonnegative and every inner
product is nonnegative by the local argument, proving the claim.
\hfill$\square$

\begin{corollary}[Exact boundary]
Start zero-state additive signed RM or RM+.  At each player event, let all
owned rows update simultaneously from one immutable exact target payoff
functional.  Then exact public feedback satisfies
\begin{equation}
\operatorname{expl}(\bar x_T,\bar y_T)
\le\frac{C_R}{2\sqrt T}+\frac{\Omega_0}{2T}.
\label{eq:app-exact-boundary}
\end{equation}
\end{corollary}

At exact feedback, Lemma~\ref{lem:app-global-monotonicity} applies to the
target game.  In Equation~\ref{eq:app-alternating-correction}, the first term
telescopes to at most $\Omega_0$ and the realized-improvement sum is
nonnegative, so its negative contribution to $C_T$ is nonpositive.  The public
and coreset debits are zero, proving the display.

The signed-RM statement uses the full uniform event-clock average.
Corollary~\ref{cor:weighted-balanced-rmplus} covers additive RM+ with the full
nondecreasing owner-event observer, including linear and quadratic weights.
The executed partial additive-RM+ quadratic-CFR+ arms instantiate the weighted
theorem together with the epoch-start output.  DCFR and PCFR+ have
recurrence-specific state transitions.

\subsection{Debit-controller quantifiers and exact separation}
\label{app:debit-controller-proof}

The controller fixes the finite game, row and public registries, global zero
initialization, nonnegative nondecreasing budget, output clock, block length,
candidate-generation rule, processor registry, outstanding cap, and uniform
row bounds before execution.  The registry designates a complete-coverage
candidate executable from every reachable block boundary, and every
selectable partial candidate has a finite history-measurable envelope.  At
completed history $H_{k-1}$ it observes the
settled account $A_{k-1}$ and pending identities $P_{k-1}$.  A partial plan is
permitted only under Equation~\ref{eq:audited-admission}; otherwise the
designated complete plan executes.  The partial plan and its predictable
envelope $\bar q_k$ are $H_{k-1}$-measurable, whereas its realized charge
$q_k$ is determined by the ensuing block and can settle later.  Complete
coverage has charge zero by
Lemma~\ref{lem:app-global-monotonicity} and requires no audit.

For block errors
$E_{\rho,k}=\sum_{t=T_{k-1}+1}^{T_k}w_t\epsilon_{\rho,t}$,
subadditivity gives
\begin{equation}
\sum_\rho\psi\!\left(\sum_{k=1}^KE_{\rho,k}\right)
\le\sum_{k=1}^K\sum_\rho\psi(E_{\rho,k}).
\label{eq:app-block-subadditivity}
\end{equation}
The payoff-decrease term is already nonnegative and additive.  Hence $Q_K$
upper-bounds both debit terms in Theorem~\ref{thm:weighted-target-debit}.  If
block $k$ is partial, its prior pending blocks together with the new block
form a set of at most $\ell$ partial blocks, and
\[
Q_k=A_{k-1}+\sum_{j\in P_k}q_j
\le b(T_k)+\sum_{j\in P_k}\bar q_j
\le b(T_k)+R_k^{(\ell)}.
\]
If block $k$ is complete, $Q_k=Q_{k-1}$, while both $b$ and
$R_k^{(\ell)}$ are nondecreasing.  Audit settlement leaves $Q_k$ unchanged.
The additive-RM+ envelope in
Lemma~\ref{lem:app-weighted-rmplus-envelope} is uniform over the registered
candidate family, proving Theorem~\ref{thm:debit-controller}.

For completeness, suppose a deterministic candidate continuation is replayed
from a copied state before selection and assigned its exact charge $q_H(a)$.
Requiring
$Q_{k-1}+q_H(a)\le b(T_k)$ and selecting the least-cost feasible candidate
gives the prospective zero-overdraft implementation used as an appendix
comparator.  Its guarantee follows from the same block subadditivity with
$Q_K\le b(T_K)$; it is not the quantifier order used by the causal controller.

\begin{proposition}[Dominance over symmetric error charging]
\label{prop:symmetric-charge-separation}
For a completed block, let $E_\rho^a$ and $A^a$ denote its realized exact
cumulative row error and payoff charge, and define
\[
q_{\rm one}(a)=\sum_\rho\psi(E_\rho^a)+A^a,
\qquad
q_{\rm sym}(a)=\sum_\rho\lVert E_\rho^a\rVert_\infty+A^a.
\]
Then $q_{\rm one}(a)\le q_{\rm sym}(a)$ for every candidate.  On any common
executed prefix, symmetric feasibility therefore implies one-sided
feasibility under the same cumulative budget.  The inclusion is strict.
\end{proposition}

\paragraph{Proof.}
The pointwise inequality $[-E_\rho^a(b)]_+\le|E_\rho^a(b)|$, maximized over
actions and summed over rows and blocks, proves the comparison.  In the
witness below, $S^+$ has errors $(3/4,3/4)$ and $(0,1)$ and zero payoff charge,
so $q_{\rm one}(S^+)=0$ while $q_{\rm sym}(S^+)=7/4$. \hfill$\square$

For the finite order witness below, exact replay gives
\begin{equation}
\begin{array}{c|cc|c|c}
 & E_0 & E_1 & \sum_td_t & q \\
\hline
S^+ & (3/4,3/4) & (0,1) & 0 & 0 \\
S^- & (-3/8,5/8) & (-3/4,3/4) & 1/4 & 11/8.
\end{array}
\label{eq:app-charge-witness}
\end{equation}
Indeed, $\psi(E_0^-)=3/8$ and $\psi(E_1^-)=3/4$, while both gauges vanish for
$S^+$.  Since all declared metadata in
Proposition~\ref{prop:app-public-order-witness} agree, any charge based only on
those summaries has one common value; soundness for $S^-$ forces that value to
be at least $11/8$.  This proves
Proposition~\ref{prop:metadata-separation}.

\begin{proposition}[Finite public-order necessity witness]
\label{prop:app-public-order-witness}
There is a finite two-player zero-sum perfect-recall EFG and two four-event
public streams with the same registry and law, two complete balanced cycles,
batch width, label multiset, callback interface, symbolic work, task-graph shape,
$D_4=1$, and $P_4=4$, but different cumulative target-feedback errors.  Their
dense pre/pre NashConv values are $1/16$ and $19/32$; their epoch-start
NashConv values are $0$ and $7/16$.
\end{proposition}

\paragraph{Proof.}
Player zero chooses row $A$ or $B$.  Player one chooses column $L$ or $R$
without observing that row, after which a uniformly random public terminal
label selects
\begin{equation}
M_1=\begin{pmatrix}3&3\\1&1\end{pmatrix},\qquad
M_2=\begin{pmatrix}-1&0\\2&-1\end{pmatrix},\qquad
\bar M=\frac{M_1+M_2}{2}
=\begin{pmatrix}1&3/2\\3/2&0\end{pmatrix}.
\label{eq:app-order-matrices}
\end{equation}
This is a perfect-recall EFG; the upstream information sets are shared across
the two terminal public labels.  Singleton scaled feedback equals the selected
conditional matrix $M_j$.  Start both signed-RM states at zero and
compare $S^+=(1,2,1,2)$ with $S^-=(2,1,2,1)$.  Exact alternating updates give
the pre/pre profiles
\begin{center}
\small
\begin{tabular}{@{}c@{\quad}cc@{\qquad}cc@{}}
\toprule
$t$ & $x_t(S^+)$ & $y_t(S^+)$ & $x_t(S^-)$ & $y_t(S^-)$\\
\midrule
1 & $(1/2,1/2)$ & $(1/2,1/2)$ & $(1/2,1/2)$ & $(1/2,1/2)$\\
2 & $(1,0)$     & $(1/2,1/2)$ & $(0,1)$     & $(0,1)$\\
3 & $(1,0)$     & $(1,0)$     & $(3/4,1/4)$ & $(0,1)$\\
4 & $(1,0)$     & $(1,0)$     & $(1,0)$     & $(0,1)$\\
\bottomrule
\end{tabular}
\end{center}
The playerwise transition tolls are $(1,0,0,1)$ and $(0,1,0,1)$ for $S^+$,
and $(1,3/2,1/2,0)$ and $(1,0,0,0)$ for $S^-$; both sum to $P_4=4$.
Both coefficient streams have internal prefix magnitude one and reset after
each pair.

Recomputing exact target rows with $\bar M$ at the same pre/pre and post/pre
origins gives
\begin{align}
(E_0^+,E_1^+)&=((3/4,3/4),(0,1)),\\
(E_0^-,E_1^-)&=((-3/8,5/8),(-3/4,3/4)).
\end{align}
The dense average profiles are
$((7/8,1/8),(3/4,1/4))$ and
$((9/16,7/16),(1/8,7/8))$.  Row maximization minus column minimization in
$\bar M$ gives gaps $1/16$ and $19/32$.  The epoch-start output retains
rounds $\{1,3\}$, giving
profiles $((3/4,1/4),(3/4,1/4))$ and
$((5/8,3/8),(1/4,3/4))$, with gaps $0$ and $7/16$.
Thus count, work, and scalar discrepancy/path metadata do not determine the
signed order--profile pairing. \hfill$\square$

\section{Experimental Details}
\label{app:registers}

\subsection{Implementation of the theorem contract}

The HUNL implementation satisfies the additive-lane contract directly.  Its
15 registered public cuts contain 48 river outcomes each and partition 720
public branches.  The 8,592 descendant nodes and 3,072 mutable decision-state
regions below those cuts are pairwise disjoint, and no registered cut descends
from another.  The 5,520 terminals below the cuts together with 14 exact
pre-river folds partition all 5,534 terminals.  Consequently every terminal
contribution belongs either to one declared public lane or to the exact
remainder, as required by Lemma~\ref{lem:app-public-cut-lanes}.

For every player event, all selected components are evaluated at one immutable
pre-update profile, centered at that same origin, aggregated, and followed by
one additive RM or RM+ update.  Unselected lane components are conceptual zero
callbacks, so owner and output clocks advance without changing learner state.
The native validation compares transcripts, regrets, current profiles,
strategy sums, averages, selected outcomes, and terminal-work fingerprints.

\begin{table}[h]
\caption{Mapping the balanced-feedback assumptions to the executed
partial RM+ runs.}
\label{tab:app-partial-instantiation}
\centering\footnotesize
\begin{tabularx}{\linewidth}{@{}p{0.19\linewidth}Xp{0.16\linewidth}@{}}
\toprule
Interface & Executed realization & Validation \\
\midrule
Additive owner lanes
& 15 nonnested cuts, 48 outcomes per cut, and one owner-local lane state per
cut and update player
& Structural pass \\
Same-origin feedback
& Pre-event action values centered at the same traversal origin; one
postorder update
& Native tests \\
Full conceptual clock
& Zero callbacks for unselected components; no owner- or output-clock
compression
& Source and replay \\
Balanced delivery
& Persistent cyclic $B\in\{8,16,24\}$ with
$m=48/B\in\{6,3,2\}$ and multiplier $48/B$
& Ledger identity \\
Learner and output
& Additive RM+, player order 0 then 1, quadratic $w_t=t^2$, and an
epoch-start coreset carrying complete epoch weight
& 110 commands \\
Exact boundary
& Complete enumeration at $m=1$ with identical coverage-clock and native
quadratic outputs
& Bit identity \\
\bottomrule
\end{tabularx}
\end{table}

\subsection{Protocol and resource accounting}
\label{app:claim-timing}

The order and width studies use disjoint seed banks.  Seeded quality
comparisons report means and deterministic paired-percentile intervals from
20,000 resamples.  Physical studies use ten sequential matched blocks with
rotated and reversed arm order; ratios are paired geometric means with
two-sided 95\% $t$ intervals on log ratios.  Every exact arm is deterministic
at fixed configuration, and every reported semantic twin reproduces the
corresponding learner and current-profile fingerprints.

Kernel CPU includes traversal, learner updates, and the method's epoch-start
output accumulation.  It excludes input loading, immutable game construction,
JSON serialization, and post-kernel exact exploitability evaluation.  The
reported runs execute on one arm64 macOS host with ten performance cores;
numerical-library thread counts are fixed at one.  The matched-CPU analysis
uses process CPU rather than wall time, so a partial run is compared with the
deepest same-block exact checkpoint whose charged CPU does not exceed the
partial run's charge.

The RM+ frontier contains 280 runs: three partial widths and exact coverage,
seven complete-epoch horizons, and ten new order seeds.  Widths $B=16$ and
$B=24$ beat the exact CPU staircase in 69 of 70 same-seed comparisons each;
20 of the 21 width-by-CPU cells satisfy the prespecified eight-of-ten
criterion.
The single loss is $B=8$ at $R=1{,}152$, where its measured
$1.353\times$ per-outcome premium permits a deeper exact run.  Repeating the
same design with signed CFR produces 21 of 21 successful cells and
partial-coverage gains of 8.4--27.2\%.  Per-outcome CPU premia for
$B=24,16,8$ are $1.146,1.197,1.353\times$ under RM+ and
$1.148,1.200,1.347\times$ under signed CFR.

\paragraph{Certificate evidence tiers.}
The certificate replay retains online sufficient statistics for the target,
delivered, discrepancy, quotient-debit, and target-payoff terms.  Exact target
rows are reconstructed from the chance-null contribution and the complete
quotient partition; the selected quotient is recovered algebraically from
ordinary delivered feedback.  This removes two redundant complete-tree
passes without changing any learner, component-history, or certificate
fingerprint.

A prospective 36-process audit separates this core from two optional evidence
layers.  Each of nine $(B,\text{seed})$ cells contains a learner-only run, the
core certificate, the core plus repeated dense-output evaluation, and the
complete path/weighted-coreset attachment.  All trajectory, identity, bound,
cross-tier, and no-swap checks pass.  Table~\ref{tab:app-certificate-resource}
reports incremental process CPU in units of its matched learner-only run and
peak resident set size (RSS) as a ratio to that run.  The prespecified periodic class requires
median and maximum added CPU at most $2.5$ and $4$ learner runs and median and
maximum RSS at most $4\times$ and $6\times$, respectively.

\begin{table}[h]
\caption{Resource cost of successively stronger certificate evidence.  CPU is
the increment over a trajectory-matched learner-only execution; RSS is the
total peak-memory ratio.}
\label{tab:app-certificate-resource}
\centering\footnotesize
\setlength{\tabcolsep}{3.0pt}
\begin{tabular}{@{}lrrrrl@{}}
\toprule
Evidence tier & Med. CPU & Max CPU & Med. RSS & Max RSS & Class \\
\midrule
Core feedback certificate & $1.56\times$ & $3.43\times$ & $3.52\times$ & $4.42\times$ & Periodic \\
$+$ dense output readout & $2.29\times$ & $4.21\times$ & $3.66\times$ & $4.58\times$ & Offline \\
$+$ path and coreset & $3.01\times$ & $4.74\times$ & $3.80\times$ & $4.72\times$ & Offline \\
\bottomrule
\end{tabular}
\end{table}

The dense readout and remaining theorem quantities add medians of $0.74$ and
$0.73$ learner executions, respectively.  Within the core shadow, omitted-
component quotient traversals account for a median $88.2\%$ of process CPU;
chance-null traversal, preparation, and validation account for $6.1\%$,
$4.0\%$, and $1.6\%$.  Thus the residual resource target is the theorem's
quotient batch rather than generic strategy caching.  The campaign's ten
concurrent single-worker processes use $95.1\%$ of their aggregate worker
budget; process CPU, not concurrent wall time, determines the classifications.

\subsection{Public order, width, and horizon}

Table~\ref{tab:temporal-balance} isolates public order at a common budget.
The three partial schedules use ten seeds; complete coverage repeats the same
deterministic endpoint.  A read-only equal-coverage observer retains 84.7\%
of the IID--reshuffle contrast.  In the separate additive-RM+ schedule
control at $B=16$, persistent cyclic feedback beats IID and fresh reshuffling
at all three horizons and wins all ten seedwise comparisons at
$R=1{,}536$.

\begin{table}[h]
\caption{Mean exploitability at 3,072 selected outcomes per lane (mbb/g;
lower is better).}
\label{tab:temporal-balance}
\centering\small
\begin{tabular}{@{}lrrrr@{}}
\toprule
Schedule & $B=1$ & $B=8$ & $B=24$ & $B=48$ \\
\midrule
IID per-event draws & 5,041.819 & 4,714.934 & 3,932.169 & 2,651.182 \\
Fresh epoch reshuffle & 2,277.814 & 2,354.077 & 2,548.787 & 2,651.182 \\
Persistent cyclic & 1,963.959 & 1,964.198 & 2,270.390 & 2,651.182 \\
\bottomrule
\end{tabular}
\end{table}

The three prespecified affine-cancelling alternatives retain the same
learner, seed-derived base permutation, selected-outcome budget, and work
ledger as the persistent control.  Cyclic order wins all 20 primary pairs.
The retained component records close every decomposition and attribute
87--97\% of the debit gap between the rotation and cyclic orders to the
order-induced change in the adaptive component path, rather than to the
position term computed on the same path.

Table~\ref{tab:app-partial-rmplus-full-grid} gives the complete held-out
shallow grid used by Section~\ref{sec:experiments}.  All 21 partial-versus-exact
paired intervals have positive lower endpoints.  These checkpoints arise
from 40 independent seed/schedule paths rather than 280 independent
experiments.  The same registered grid on Turn Subgame~1 again produces 21
positive contrasts: 19 cells have ten of ten seedwise wins, one has nine, and
one has eight.  All three terminal-width cells have ten of ten wins, with
relative reductions of 19.6\%, 14.1\%, and 5.3\% for $B=8,16,24$.

\begin{table}[h]
\caption{Held-out additive-RM+ epoch-coreset exploitability in mbb/g.  Entries
are mean (sample standard deviation) over ten public-order seeds; exact
enumeration is deterministic.}
\label{tab:app-partial-rmplus-full-grid}
\centering\small
\begin{tabular}{@{}rrrrr@{}}
\toprule
$R$/lane & $B=8$ & $B=16$ & $B=24$ & Exact $B=48$ \\
\midrule
384   & 9,321.574 (334.162) & 10,437.822 (301.417) & 11,545.746 (770.192) & 13,189.670 (0) \\
576   & 5,212.464 (229.020) &  5,960.795 (247.145) &  7,050.925 (456.458) &  8,108.295 (0) \\
768   & 3,403.397 (134.937) &  3,716.135 (164.177) &  4,550.150 (297.309) &  5,595.911 (0) \\
960   & 2,548.966 (113.726) &  2,554.505 (102.718) &  3,039.417 (196.017) &  3,914.246 (0) \\
1,152 & 2,080.202 ( 99.266) &  1,959.033 ( 80.512) &  2,165.101 (129.958) &  2,876.704 (0) \\
1,344 & 1,795.931 ( 99.888) &  1,614.202 ( 69.950) &  1,672.635 ( 92.609) &  2,229.986 (0) \\
1,536 & 1,599.968 (115.341) &  1,391.091 ( 70.942) &  1,382.963 ( 81.612) &  1,814.056 (0) \\
\bottomrule
\end{tabular}
\end{table}

The depth ladder reuses the same widths and extends the outcome budget to 512
exact-equivalent rounds.  Every partial width crosses complete coverage
between rounds 32 and 64.  At round 512, exact coverage reaches 31.1 mbb/g,
whereas $B=8,16,24$ reach 273, 233, and 171 mbb/g.  The complete and partial
paths remain semantically attached to their declared output clocks throughout
the ladder.

\subsection{Target-valid time to exact quality}
\label{app:target-valid-time}

A separate preregistered campaign compares three additive-RM+ arms under the
same quadratic output clock: exact enumeration at $B=48$, a persistent cyclic
schedule at $B=16$, and conditionally target-valid IID width-$16$ public
sampling.  Exact quality is checked after every 48 selected outcomes per lane,
and each arm may consume at most 3,072 outcomes.  The three target values are
the exact-CFR+ exploitabilities at 960, 1,152, and 1,536 outcomes per lane.
Ten new public-order seeds define paired, counterbalanced blocks.  Each solve
uses one physical worker; ten solves execute concurrently on the host's ten
performance cores.  All trajectory, output-clock, target-readout, source, and
no-swap checks pass, and every arm reaches all three targets.

\begin{table}[h]
\caption{Time to three exact HUNL quality targets.  Outcome entries are means
over ten cells.  CPU ratios are paired geometric means with two-sided 95\%
intervals; ratios above one favor persistent $B=16$.}
\label{tab:app-target-valid-time}
\centering\footnotesize
\setlength{\tabcolsep}{2.8pt}
\begin{tabular}{@{}lrrrl@{}}
\toprule
\shortstack{Target\\(mbb/g)} & Outcomes (exact/cyclic/IID) & Exact/cyclic CPU & IID/cyclic CPU &
Exact/cyclic charged \\
\midrule
$3,914.246$ & 960 / 763 / 1,459
& $1.055$ [$0.958,1.163$] & $1.918$ [$1.780,2.066$]
& $0.914$ [$0.827,1.010$] \\
$2,876.704$ & 1,152 / 902 / 1,790
& $1.045$ [$0.956,1.143$] & $2.007$ [$1.752,2.300$]
& $0.905$ [$0.827,0.991$] \\
$1,814.056$ & 1,536 / 1,238 / 2,534
& $1.004$ [$0.880,1.145$] & $1.927$ [$1.648,2.252$]
& $0.866$ [$0.771,0.973$] \\
\bottomrule
\end{tabular}
\end{table}

The persistent arm uses 19--22\% fewer outcomes than exact CFR+, but each
partial outcome costs 16--25\% more learner CPU because three partial events
replace one full-cut event and repeat fixed prefix, policy, and learner work.
These effects nearly cancel, producing exact-CFR+ learner-time parity.  IID
requires approximately twice as many outcomes and learner CPU as persistence,
which isolates the finite-resource value of deterministic coverage structure
relative to the conditionally target-valid IID control.

The final table column additionally charges the external exact-output
evaluation at each stopping checkpoint.  Exact full-cut traversal has already
traversed the evaluator's complete support, whereas partial traversal has not;
the resulting point estimates favor exact CFR+ by 9--15\%, with the
medium-target interval including one.  The core feedback
certificate is a periodic audit rather than part of the online stopping path.
At these horizons it adds a median 1.58--1.72 matched learner executions, so
its resource class is reported separately in
Table~\ref{tab:app-certificate-resource}.

\subsection{Independent public-Leduc target-transfer study}
\label{app:leduc-transfer}

The second-domain study was fixed before the held-out paths were opened.  It
uses the ten-card public-Leduc game with five ranks, two copies per rank, two
betting streets, and one public community-card reveal.  The additive registry
contains the ten physical card identities.  A component is zero when its card
is privately held, and fold payoffs before the reveal are divided uniformly
among the labels.  Summing all components therefore reconstructs the separate
dense traversal.  The learner is simultaneous signed CFR: both players consume
feedback at the pre-round profile, so $L_{\rm alt}=0$, and the reported output
is the standard own-reach-weighted average of those profiles.

Partial widths $B\in\{2,5\}$ use either one repeated random permutation or a
fresh permutation at every epoch; complete coverage uses $B=10$.  The ten
held-out seeds are 1103, 2111, 3109, 4111, 5101, 6101, 7103, 8101, 9103, and
10103.  At each $Q\in\{1,2,4,8,16,32,64\}$, every arm has evaluated exactly
$10Q$ components, while a partial path has made $10Q/B$ joined learner
transitions.  The resulting register contains 40 partial paths, one complete
component path, and one independent dense-reference path.  The fixed protocol
compares quality at matched component budgets.

\begin{table}[h]
\caption{Public-Leduc NashConv (lower is better) at equal selected-component
budgets.  The partial median and win count summarize 40 held-out paths; the
complete path is deterministic.}
\label{tab:app-leduc-budget}
\centering\small
\begin{tabular}{@{}rrrr@{}}
\toprule
Coverage epochs $Q$ & Complete & Partial median & Partial wins \\
\midrule
1  & 7.86250 & 5.96111 & 40/40 \\
2  & 4.12292 & 4.53539 &  4/40 \\
4  & 2.71910 & 2.74521 & 18/40 \\
8  & 1.76906 & 1.51510 & 40/40 \\
16 & 1.16519 & 1.01823 & 40/40 \\
32 & 0.73236 & 0.62605 & 40/40 \\
64 & 0.50194 & 0.38240 & 40/40 \\
\bottomrule
\end{tabular}
\end{table}

\begin{table}[h]
\caption{Public-Leduc validity and order checks.  Residual thresholds were
fixed at $10^{-9}$; order thresholds were 15/20 overall and 7/10 per width.}
\label{tab:app-leduc-validity}
\centering\footnotesize
\begin{tabularx}{\linewidth}{@{}lXr@{}}
\toprule
Check & Observed result & Status \\
\midrule
Dense/component parity
& All 7 checkpoints; max NashConv difference $2.0\times10^{-15}$ and
identical quantized learner/profile fingerprints & pass \\
Target and Abel identities
& Maximum residuals $5.56\times10^{-13}$ and $9.99\times10^{-15}$ & pass \\
Direct target certificate
& 287/287 component-path checkpoints; minimum slack $0.1995$ & pass \\
Delivered-plus-debit certificate
& 287/287 checkpoints; minimum slack $0.2335$ & pass \\
Complete-coverage debit
& Exactly zero at 7/7 checkpoints & pass \\
Fixed-cycle quality comparisons
& 12/20 overall; 6/10 at each width & criterion unmet \\
Fixed-cycle debit comparisons
& 13/20 overall; 6/10 at $B=2$ and 7/10 at $B=5$ & criterion unmet \\
Debit--quality sign agreement
& 13/20 pairs & criterion unmet \\
\bottomrule
\end{tabularx}
\end{table}

The two-sided exact sign-test values are $0.503$ for quality and $0.263$ for
normalized debit, with no ties.  The prespecified universal fixed-cycle
selector is therefore unsupported.  All six target-transfer validity checks
pass, and every partial and complete path is certified against the dense game.
At $Q=64$, fixed-cycle and fresh-reshuffle medians are
$0.34897$ and $0.35284$ for $B=2$, and $0.40603$ and $0.40611$ for $B=5$.
Debit signs agree with the NashConv differences in 13 of 20 pairs, while
delivered-regret differences agree in 18 of 20 post hoc pairs.  The latter
therefore carries additional information for finite-quality schedule ranking.

\subsection{Causal debit census and controller studies}
\label{app:debit-controller-experiment}

The causal census uses alternating signed RM, ten-round common blocks, widths
$2,5,10$, and development seeds 1201, 2203, 3203, and 4201.  Three fixed-width
reference paths per seed produce 48 completed-history states.  All three
candidate widths are replayed from each state, yielding 144 records.  Complete
width has zero debit throughout; partial debit ranges from 2.733 to 53.439.
All three widths occur on the exact work--debit Pareto frontier.

The symmetric-charge replay retains the same 48 states and 144 continuations.
For every row error $E_\rho$, it compares $\psi(E_\rho)$ with
$\lVert E_\rho\rVert_\infty$.  All source charges and successor fingerprints
reproduce.  The inequality is strict on all 96 partial records, with median
symmetric-to-one-sided ratio $2.325$.  Table~\ref{tab:app-symmetric-charge}
reports the width-resolved result at the standalone ten-round budget
$12\sqrt{10}$; both charges remain zero under complete coverage.

\begin{table}[h]
\caption{One-sided versus symmetric charge on the 48 causal contexts.
Admissions use the fixed standalone block budget $12\sqrt{10}$.}
\label{tab:app-symmetric-charge}
\centering\small
\begin{tabular}{@{}rrrrr@{}}
\toprule
Width & One-sided median & Symmetric median & One-sided admitted & Symmetric admitted \\
\midrule
2  & 22.151 & 49.121 & 36/48 & 8/48 \\
5  &  7.180 & 16.301 & 48/48 & 48/48 \\
10 &  0     &  0     & 48/48 & 48/48 \\
\bottomrule
\end{tabular}
\end{table}

\paragraph{Lagged causal controller.}
With $\kappa=12$ fixed, a four-order development tournament compared fixed
widths 2 and 5, a fixed alternation, the lagged completed-history rule from
Section~\ref{sec:method}, and exact coverage.  The lagged rule lay on the
development work--quality Pareto set and was selected before a disjoint
ten-order evaluation.  It executed each selected block once and settled its
exact charge before the next decision.  All overdraft, zero-charge,
matched-work, and fingerprint checks pass.  Table~\ref{tab:app-causal-controller}
reports the complete held-out register; $X$ denotes exact width 10.

\begin{table}[h]
\caption{Held-out lagged causal DB-CFR register.  Gain is best
matched-recurrence-work fixed-width NashConv minus controller NashConv.}
\label{tab:app-causal-controller}
\centering\scriptsize
\setlength{\tabcolsep}{2.8pt}
\begin{tabular}{@{}rcrrrr@{}}
\toprule
Seed & Width string & Work & DB-CFR & Best fixed & Gain \\
\midrule
47017 & 522522222552 & 720 & 0.289511 & 0.334781 & 0.045270 \\
48017 & 525522222525 & 780 & 0.295285 & 0.360470 & 0.065185 \\
49003 & 522222222X22 & 700 & 0.318324 & 0.384413 & 0.066089 \\
50021 & 525555222222 & 780 & 0.299940 & 0.322948 & 0.023008 \\
51001 & 525522222522 & 720 & 0.333773 & 0.367134 & 0.033362 \\
52009 & 522255552222 & 780 & 0.289810 & 0.332691 & 0.042881 \\
53003 & 522522555222 & 780 & 0.286374 & 0.313397 & 0.027023 \\
54001 & 522222255522 & 720 & 0.306304 & 0.352147 & 0.045843 \\
55009 & 522252252222 & 660 & 0.312263 & 0.398279 & 0.086016 \\
56003 & 525522222225 & 720 & 0.289894 & 0.374494 & 0.084600 \\
\bottomrule
\end{tabular}
\end{table}

The causal rule wins all ten pairs, has median gain 0.04556, uses all three
registered widths, and has median installed recurrence work 720 components.
It evaluates no unselected continuation.  Relative to cost-ordered
prospective lookahead, its median candidate-replay ratio is 0.728.  The exact
post-block observer is recorded separately from installed recurrence work.

\paragraph{Prospective comparator.}
The development controller evaluates
$\kappa\in\{0,4,8,12,16\}$ for
$b(T)=\kappa\sqrt T$.  Candidate selection receives only width, selected work,
exact charge, and checkpoint identity.  The prespecified selection rule promotes a
value only if its dynamic path beats the best fixed width at no greater
selected work on at least three of four orders and has positive paired median
improvement.  Only $\kappa=12$ passes, with four wins and median improvement
0.0666.  The prospective protocol then fixes this value, eight blocks, the
matched-work comparator, and eight previously unused orders.

\begin{table}[h]
\caption{Complete prospective DB-CFR register.  Improvement is best
matched-work fixed-width NashConv minus controller NashConv.}
\label{tab:app-prospective-controller}
\centering\footnotesize
\setlength{\tabcolsep}{3pt}
\begin{tabular}{@{}rcrrrr@{}}
\toprule
Seed & Executed widths & Work & DB-CFR & Best fixed & Improvement \\
\midrule
5309  & 5,2,2,5,2,5,2,5 & 560 & 0.492083 & 0.451720 & -0.040363 \\
6311  & 5,2,2,2,5,2,5,5 & 560 & 0.450423 & 0.485398 &  0.034975 \\
7307  & 5,2,2,2,2,2,5,2 & 440 & 0.365417 & 0.562647 &  0.197230 \\
8311  & 5,2,2,2,2,5,5,2 & 500 & 0.412053 & 0.465212 &  0.053159 \\
9311  & 2,5,2,2,2,5,5,2 & 500 & 0.579293 & 0.449471 & -0.129822 \\
10301 & 5,2,2,2,2,5,2,2 & 440 & 0.437389 & 0.490160 &  0.052770 \\
11311 & 5,2,2,2,5,2,5,2 & 500 & 0.414273 & 0.483224 &  0.068950 \\
12301 & 5,2,2,5,2,2,5,5 & 560 & 0.475374 & 0.509573 &  0.034199 \\
\bottomrule
\end{tabular}
\end{table}

All eight budgets, least-work decisions, exact-fallback charges, and replay
fingerprints pass.  The prespecified transfer criterion is met with six wins,
pooled use
of both partial widths, and paired median improvement 0.043872.  Seeds 5309
and 9311 are retained negative cells.

A separate offer-search audit uses ten new public orders and compares eager
evaluation of all three continuations with cost-ordered exact certification.
Across 80 block decisions, selected widths, charges, qualities, and
learner/profile fingerprints match exactly.  Eager search evaluates 240
continuations; lazy search evaluates 107, comprising 80 selected successors
and 27 rejected cheaper offers.  Per seed, the lazy/eager replay ratio has
median $0.458$ and maximum $0.500$.  Replay count is the registered resource
measure; local process CPU is diagnostic because the host does not admit a
timing claim.  This audit does not price a continuation-valid approximation
to the selected candidate's exact semantic debit.

\subsection{Trace certificates}
\label{app:certificate-instantiation}

The signed-CFR certificate joins widths $16/24/48$ (61 paths and 2,496
checkpoints) with an independently collected width-8 bank (30 paths and 480
checkpoints).  At all 2,976 checkpoints, the target-regret assembly identity
closes to relative error at most $10^{-9}$, target regret is no larger than
delivered regret plus quotient debit, and quotient debit is exactly zero at
complete coverage.  The median certificate decreases at every one of the 374
adjacent horizon steps.

\begin{table}[h]
\caption{Uniform-clock certificate on persistent cyclic signed-CFR paths
(medians over ten seeds; chips, where 1 chip $=10$ mbb/g).  The surrogate
uses delivered regret and quotient debit; the direct bound uses measured
target regret.}
\label{tab:certificate-audit}
\centering
\footnotesize
\setlength{\tabcolsep}{2.4pt}
\begin{tabular}{@{}rrrrrrrrr@{}}
\toprule
$B$ & $T$ & $R$/lane & $\widehat{\mathcal R}_T$ & $\mathcal E_{Q,T}^\infty$ &
surrogate & bound & certified & bound/cert. \\
\midrule
8 & 192 & 1,536 & 521,934 & 73,539 & 1,655 & 984 & 355 & 2.77 \\
16 & 96 & 1,536 & 269,503 & 27,188 & 1,754 & 1,248 & 399 & 3.13 \\
24 & 96 & 2,304 & 202,590 & 15,068 & 1,342 & 1,063 & 290 & 3.66 \\
48 & 96 & 4,608 & 135,867 & 0 & 916 & 916 & 178 & 5.15 \\
\bottomrule
\end{tabular}
\end{table}

A separate 31-path replay measures the certified dense pre/pre average.  All
1,216 complete-epoch checkpoints lie below both the delivered-plus-debit
surrogate and the direct target-regret bound.  The per-width direct-bound
ratios remain nearly horizon independent, ranging from
$2.65$--$2.95\times$ at $B=8$ to $4.8$--$5.2\times$ at complete coverage.

\paragraph{Weighted RM+ certificate.}
\phantomsection\label{app:rmplus-certificate}
RM+ clips its learner state, so a read-only accumulator records the unclamped
delivered sum at the same update point.  Thirty trajectory-identical replays
cover widths $8/16/24$, ten held-out seeds, and seven complete-epoch horizons.
The delivered-sum identity closes to $7.1\times10^{-14}$, and the direct
weighted target-regret bound covers the weighted dense average at all 210
checkpoints.

The reported epoch-start output is a compression of that certified dense
average.  Its cell-median exploitability is $1.04$--$1.28\times$ the certified
profile's exploitability.  The exact weighted endpoint and within-epoch
variation terms produce $\mathcal E_{Q,T}^{w,\infty}$, which covers the direct
delivered-minus-target row error at every checkpoint.  Delivered weighted
positive local regret plus this debit covers weighted target positive local
regret throughout.  Two exact signed best responses then instantiate
Equation~\ref{eq:app-signed-coreset-supports}; their correction separately
covers the measured epoch-start-minus-dense increase at every row.

Consequently, Equation~\ref{eq:app-complete-weighted-output-certificate}
covers all 210 reported outputs, and every final bound is below the
payoff-span exploitability bound.  Across the 21 width--horizon cells, median
final-bound ratios are $11.11$--$30.20\times$, compared with
$3.68$--$9.60\times$ for the direct-target diagnostic.  The quotient term
contributes a median $65.0\%$ of the delivered-plus-debit dense bound, whereas
the signed-support correction contributes a median $2.9\%$ of the final
bound.  The retained ledger therefore both completes the reported-output
certificate and locates its principal slack in the upstream component and
positive-regret transfer.

\section{Reproducibility}
\label{app:repro}

The accompanying package contains the native solvers, theorem verifiers,
campaign protocols, command manifests, raw transcripts, summary artifacts,
and tests.  Every figure and table is linked to retained per-value provenance,
and all principal trajectory comparisons carry learner, profile, input, and
configuration fingerprints.

The quality evidence covers two released full-range HUNL turn inputs and an
independently implemented public-Leduc game.  The public-Leduc package retains
the fixed protocol, prediction seal, 42 path records, 294 total checkpoints,
and dense-reference controls; 287 component-path checkpoints
enter the certificate audit.  The matched-CPU evidence uses a host with ten
performance cores and the kernel-CPU accounting described above.  Source,
input, and artifact digests are retained with the campaigns; timing statements
are restricted to the recorded execution environment and charged interval.

\end{document}